\documentclass[twocolumn]{aastex631}
\usepackage{graphicx}
\usepackage{amsmath}
\usepackage{epsfig}
\usepackage{enumitem}
\usepackage{hyperref}
\usepackage{xcolor}
\usepackage{nicefrac}
\usepackage{threeparttable}
\usepackage{multirow}

\begin{document}

\title{Long-rising Type II Supernovae in the Zwicky Transient Facility Census of the Local Universe}

\correspondingauthor{Tawny Sit}
\email{sit.6@osu.edu}

\author[0000-0001-8208-9755]{Tawny Sit}
\affiliation{Department of Astronomy, The Ohio State University, Columbus, OH}
\affiliation{Cahill Center for Astrophysics, California Institute of Technology, Pasadena CA 91125, USA}

\author{Mansi M. Kasliwal}
\affiliation{Cahill Center for Astrophysics, California Institute of Technology, Pasadena CA 91125, USA}

\author{Anastasios Tzanidakis}
\affiliation{Department of Astronomy, University of Washington, Seattle, WA 98195, USA}

\author{Kishalay De}
\affiliation{MIT-Kavli Institute for Astrophysics and Space Research, Cambridge, MA 02139, USA}

\author{Christoffer Fremling}
\affiliation{Cahill Center for Astrophysics, California Institute of Technology, Pasadena CA 91125, USA}

\author[0000-0003-1546-6615]{Jesper Sollerman}
\affiliation{The Oskar Klein Centre, Department of Astronomy, Stockholm University, AlbaNova, SE-106 91 Stockholm, Sweden}

\author[0000-0002-3653-5598]{Avishay Gal-Yam}
\affiliation{Department of Particle Physics and Astrophysics, Weizmann Institute of Science, 76100 Rehovot, Israel}

\author{Adam A. Miller}
\affiliation{Center for Interdisciplinary Exploration and Research in Astrophysics (CIERA), Department of Physics and Astronomy, Northwestern University, Evanston, IL 60208, USA}

\author{Scott Adams}
\affiliation{Cahill Center for Astrophysics, California Institute of Technology, Pasadena CA 91125, USA}

\author{Robert Aloisi}
\affiliation{Department of Astronomy, University of Wisconsin-Madison, 475 North Charter Street, Madison, WI 53706, USA}

\author{Igor Andreoni}
\affiliation{Joint Space-Science Institute, University of Maryland, College Park, MD 20742, USA}
\affiliation{Department of Astronomy, University of Maryland, College Park, MD 20742, USA}
\affiliation{Astrophysics Science Division, NASA Goddard Space Flight Center, Greenbelt, MD 20771, USA}

\author{Matthew Chu}
\affiliation{Cahill Center for Astrophysics, California Institute of Technology, Pasadena CA 91125, USA}

\author{David Cook}
\affiliation{IPAC, California Institute of Technology, 1200 E. California Blvd, Pasadena, CA 91125, USA}

\author{Kaustav Kashyap Das}
\affiliation{Cahill Center for Astrophysics, California Institute of Technology, Pasadena CA 91125, USA}

\author{Alison Dugas}
\affiliation{Institute for Astronomy, University of Hawai’i, Honolulu, HI 96822, USA}

\author[0000-0001-5668-3507]{Steven L. Groom}
\affiliation{IPAC, California Institute of Technology, 1200 E. California Blvd, Pasadena, CA 91125, USA}

\author[0000-0002-9017-3567]{Anna Y. Q. Ho}
\affiliation{Department of Astronomy, Cornell University, Ithaca, NY 14853, USA}

\author{Viraj Karambelkar}
\affiliation{Cahill Center for Astrophysics, California Institute of Technology, Pasadena CA 91125, USA}

\author[0000-0002-0466-1119]{James D. Neill}
\affiliation{Cahill Center for Astrophysics, California Institute of Technology, Pasadena CA 91125, USA}

\author[0000-0002-8532-9395]{Frank J. Masci}
\affiliation{IPAC, California Institute of Technology, 1200 E. California Blvd, Pasadena, CA 91125, USA}

\author[0000-0002-7226-0659]{Michael S. Medford}
\affiliation{Department of Astronomy, University of California, Berkeley, Berkeley, CA 94720}
\affiliation{Lawrence Berkeley National Laboratory, 1 Cyclotron Rd., Berkeley, CA 94720}

\author{Josiah Purdum}
\affiliation{Caltech Optical Observatories, California Institute of Technology, Pasadena, CA 91125, USA}

\author{Yashvi Sharma}
\affiliation{Cahill Center for Astrophysics, California Institute of Technology, Pasadena CA 91125, USA}

\author[0000-0001-7062-9726]{Roger Smith}               
\affiliation{Caltech Optical Observatories, California Institute of Technology, Pasadena, CA 91125, USA}

\author{Robert Stein}
\affiliation{Cahill Center for Astrophysics, California Institute of Technology, Pasadena CA 91125, USA}

\author{Lin Yan}
\affiliation{Caltech Optical Observatories, California Institute of Technology, Pasadena, CA 91125, USA}

\author{Yuhan Yao}
\affiliation{Cahill Center for Astrophysics, California Institute of Technology, Pasadena CA 91125, USA}

\author{Chaoran Zhang}
\affiliation{Center for Gravitation, Cosmology, and Astrophysics, Department of Physics, University of Wisconsin, Milwaukee, WI 53201, USA}

\begin{abstract}
    SN 1987A was an unusual hydrogen-rich core-collapse supernova originating from a blue supergiant star. Similar blue supergiant explosions remain a small family of events, and are broadly characterized by their long rises to peak. The Zwicky Transient Facility (ZTF) Census of the Local Universe (CLU) experiment aims to construct a spectroscopically complete sample of transients occurring in galaxies from the CLU galaxy catalog. We identify 13 long-rising ($>$40 days) Type II supernovae from the volume-limited CLU experiment during a 3.5 year period from June 2018 to December 2021, approximately doubling the previously known number of these events. We present photometric and spectroscopic data of these 13 events, finding peak $r$-band absolute magnitudes ranging from $-15.6$ to $-17.5$ mag and the tentative detection of \ion{Ba}{2} lines in 9 events. Using our CLU sample of events, we derive a long-rising Type II supernova rate of $1.37^{+0.26}_{-0.30}\times10^{-6}$ Mpc$^{-3}$ yr$^{-1}$, $\approx$1.4\% of the total core-collapse supernova rate. This is the first volumetric rate of these events estimated from a large, systematic, volume-limited experiment.
    
\end{abstract}
\section{Introduction}

Supernova (SN) 1987A was the most recent naked-eye and most nearby supernova observed in the past century. Due to its proximity, having occurred in the Large Magellanic Cloud (LMC), SN 1987A was extremely well studied, with both neutrino and electromagnetic detections \citep{Arnett1989}. Its progenitor star was identified as the blue supergiant (BSG) star Sanduleak -69$^\circ$202 \citep{Gilmozzi1987}. Studies of core-collapse supernovae (CCSNe) prior to SN 1987A had suggested that hydrogen-rich CCSNe, known as Type II SNe (SNe II), originated from red supergiant (RSG) stars; BSG stars were not expected to explode as SNe by stellar evolution models at the time. After the explosion of Sanduleak -69$^\circ$202, single-star scenarios involving low metallicities (such as that of the LMC) or fast rotation, and binary interaction scenarios have been proposed as theoretical models for BSG explosions (see \citealt{Podsiadlowski1992} for a review). In the case of SN 1987A specifically, a binary companion could explain the unusual shape of the surrounding circumstellar medium \citep{Morris2009,Podsiadlowski1992_review}.

SN 1987A had an unusually long-rising light curve of $\sim$80 days. This long rise is broadly considered characteristic of 1987A-like SNe because it is well-explained by a compact BSG progenitor---much of the initial explosion energy is spent adiabatically expanding the compact star \citep{Arnett1989,Taddia2013,Kleiser2011}. Other long-rising SNe II found since 1987 include: SN 1998A \citep{Pastorello2005}; SN 2000cb \citep{Kleiser2011}; SN 2004ek, SN 2004em, and SN 2005ci \citep{Taddia2016}; SN 2006V and SN 2006au \citep{Taddia2012}; SN 2009E \citep{Pastorello2012}; SN 2009mw \citep{Takats2016}; PTF09gpn, PTF12gcx, and PTF12kso \citep{Taddia2016}; SN 2018hna \citep{Singh2019,Xiang2023}; SN Refsdal \citep{Rodney2016}; OGLE-2014-SN-073 \citep{Terreran2017}; and DES16C3cje \citep{Gutierrez2020}. Additionally, SN 1909A and 1982F are also 1987A-like events \citep{Pastorello2012,Taddia2013}, but have limited data coverage. Studies of long-rising SN II events suggest that, in general, they originate from more compact ($R<100$ R$_{\odot}$) and massive ($\sim$20 M$_{\odot}$) stars, have higher explosion energies ($\gtrsim$10$^{51}$ erg), and synthesize more $^{56}$Ni ($\approx$0.1 M$_{\odot}$, as opposed to $\approx$0.04 M$_\odot$ for normal SN IIP \citep{Muller2017}) \citep{Pastorello2012,Taddia2016}. There is some evidence that long-rising light curves similar to that of SN 1987A can also be produced by a non-BSG star that synthesizes substantial amounts of $^{56}$Ni such as SN 2004ek, which \citet{Taddia2016} considers an ``intermediate" event between SN 1987A and typical SNe IIP. A recent comparative study showed that these SN 1987A-like events encompass a wide range of parameters, similar to other subclasses of SNe, despite their small sample size: explosion energies ranging from $0.5-15 \times 10^{51}$erg, progenitor radii ranging from $0.2-100 \times 10^{12}$ cm, and ejecta masses ranging from $5-55 M_\odot$ \citep{Pumo2023}.

The rate of long-rising SN II events may be indicative of the variety of mechanisms that can allow a BSG to explode as a SN and massive star evolution in general (see \citealt[\S II.B.2]{Woosley2002}, and references therein). For example, a restricted-convection stellar evolution model can limit the mass range of stars that can theoretically explode as BSG to $\sim$15-20 M$_\odot$ at LMC metallicities (e.g., \citealt{Langer1991,Woosley2002}), thus limiting the number of BSG explosions possible in a given stellar population. \citet{Podsiadlowski1992} estimates that $\sim$5\% of all massive stars can die as a BSG due to a merger event, and \citet{Podsiadlowski1992_review} notes that such merger models are theoretically insensitive to metallicity. This contrasts \citet{Taddia2013}'s empirical result showing that long-rising SNe II do generally tend to occur in lower metallicity environments than typical SNe II. Previous observational estimates place the rate of SN 1987A-like events at $\approx$1-3\% of all CCSNe \citep{Taddia2016,Pastorello2012,Kleiser2011,Smartt2009}. However, these rate estimates are typically limited by the small number, diversity in object properties, and heterogeneity of the surveys that discovered them. A more systematic approach is necessary to understand long-rising SNe II as a class and their range of properties.

A systematic approach to long-rising SNe II will also help better define their luminosity function, which in turn will allow us to further constrain the volumetric rates of these events. The $r$-band absolute magnitudes of known long-rising SNe II range from about $-15.8$ to $-18.1$ mag with a median of $M_r\approx-16.3$; the majority (including SN 1987A itself) are fairly subluminous compared to normal SN IIP. These low luminosities and the need for reasonably well-monitored coverage of light curves during the rise make identification of 1987A-like SNe difficult, especially at farther distances---\citet{Pastorello2012} notes that incompleteness is significant at distances beyond $\approx$100 Mpc (distance modulus $\mu\approx35$ mag). 

Motivated thus to develop a systematic sample of long-rising SNe II, we need a sufficiently deep optical sky survey and a program to classify transients in the local universe that is not biased against low-luminosity objects. The Zwicky Transient Facility (ZTF; \citealt{Bellm2018,Graham2019}) achieves the first objective: the wide-field optical survey scans the northern sky every $\approx$3 days to a median limiting magnitude of $\approx$20.5 mag in $r$. The Census of the Local Universe (CLU) experiment of ZTF achieves the second, systematically building up a sample of all transients within a limited volume by cross-matching extragalactic transients detected by ZTF to nearby galaxies with known redshifts and classifying those transients regardless of their brightness. The CLU experiment has been used to systematically find and analyze a sample of calcium-rich gap transients \citep{De2020} and to constrain the faintest end of the SN II luminosity function (Tzanidakis et al., in prep.). As long-rising SNe II frequently have lower luminosities, similar to those of the Ca-rich gap transients (which peak at an average of $M_r\approx-16$, \citealt{De2020}), the CLU experiment is well-suited for a systematic approach to studying these unique events. 

In this paper, we present 13 long-rising Type II SNe found via the ZTF CLU experiment. We briefly describe the survey, experiment design, and sample selection in \S\ref{sec:sample selection}.
In \S\ref{sec:photometry}, we present and discuss the photometric properties of the sample and compare to the literature. We do the same for the spectroscopic properties in \S\ref{sec:spectroscopy}. A brief analysis of the host galaxies for the sample is presented in \S\ref{sec:hosts}. We estimate upper and lower limits of the volumetric rate of long-rising Type II SNe out to $\approx$200 Mpc in \S\ref{sec:rates}, and compare our rates to previous studies. Finally, in \S\ref{sec:discussion} we summarize our findings and discuss implications for future work.

\section{Sample Selection}\label{sec:sample selection}
\subsection{Census of the Local Universe}\label{sec:ZTF CLU}
The Zwicky Transient Facility (ZTF) is an optical time-domain survey on the 48-inch Schmidt telescope (P48) at Palomar Observatory \citep{Bellm2018,Graham2019,Dekany2020,ZTFdatacite}. The ZTF camera has a 47 deg$^2$ field of view, can achieve a survey speed of $\approx$3750 deg$^2$ hr$^{-1}$, and reaches a median limiting magnitude of $r_{ZTF}\approx20.5$ mag in its 30s exposures. This allows ZTF to observe the entire northern sky with $\sim$3 day cadence in $g_{ZTF}$ and $r_{ZTF}$ (hereafter $g$ and $r$-band, respectively) in the public Northern Sky Survey. Additional programs with different cadences are described in more detail in \citet{Bellm2019}.

The ZTF Census of the Local Universe (CLU) experiment aims to construct a spectroscopically complete sample of transients in the local universe. During ZTF Phase I, CLU was a volume-limited supernova survey where sources at less than 200 Mpc ($z\lessapprox0.05$) and spatially coincident (within 100$\arcsec$) with a galaxy in the Census of the Local Universe (CLU) galaxy catalog \citep{Cook2019} were assigned spectroscopic follow-up for classification. The CLU galaxy catalog contains $\approx$234,500 galaxies with known distances, compiled from preexisting spectroscopic galaxy surveys and the CLU-H$\alpha$ survey (for further details, see \citealt{Cook2019}). For ZTF Phase II, starting in January 2021, the CLU experiment was redesigned as a volume-luminosity limited supernova survey. The selection criteria for spectroscopic follow-up was tightened to $<$150 Mpc, within 30 kpc of or visibly associated with a CLU galaxy, and peaking below $-17$ in absolute magnitude.

To select transients for the CLU experiment sample, ZTF difference-image alerts \citep{Masci2018,Patterson2018} are first passed through a custom filter implemented on the GROWTH marshal (for ZTF-I; \citealt{Kasliwal2019}) or the Fritz portal (for ZTF-II; \citealt{Duev2019,vanderWalt2019}). This filter cross-matches alerts to the CLU galaxy catalog and makes several additional cuts to remove stars, solar system objects, and bogus sources. Of the $\sim$10$^6$ alerts produced by ZTF on a typical night, $\sim$200 pass this automatic filter. Transients that pass the filter are then vetted by human scanners to further remove variable stars, active galactic nuclei (AGN), and other non-SN sources prior to being assigned spectroscopic follow-up based on the criteria described previously. For more detailed discussion of the CLU experiment selection process, see \citet{De2020}.

\begin{deluxetable*}{ccccc}[!ht]
\tablewidth{0pt}
\tablecaption{ZTF CLU Sample of 13 Long-rising SNe II \label{table: SN info}}
\tablehead{
\colhead{} & \colhead{$\alpha$ (J2000)}\vspace{-0.4cm} & \colhead{$\delta$ (J2000)} & \colhead{} & \colhead{} \\
\colhead{SN Name / ZTF ID} & \colhead{}\vspace{-0.4cm} & \colhead{} & \colhead{Host Galaxy} & \colhead{Redshift\tablenotemark{a}}\\
\colhead{} & \colhead{(hh:mm:ss)} & \colhead{(dd:mm:ss)} & \colhead{} & \colhead{}
}
\startdata
SN 2018cub / ZTF18aaikcbb & 15:05:42.45 & +60:47:43.99 & WISEA J150541.98+604751.4 & 0.044 \\
SN 2018lrq / ZTF18aaxzlmy & 13:34:51.39 & +34:03:20.23 & MCG+06-30-045 & 0.025 \\
SN 2018ego / ZTF18ablhrpz & 15:52:54.03 & +19:58:05.41 & 2MASX J15525218+1958107 & 0.037 \\
SN 2018lsg / ZTF18abtjmns & 20:46:44.61 & -01:22:07.90 & CGCG 374-012 & 0.025 \\
SN 2018imj / ZTF18abytyif & 06:51:05.75 & +12:54:55.78 & IC 0454 & 0.013 \\
SN 2018hna / ZTF18acbwaxk & 12:26:12.08 & +58:18:50.88 & UGC 07534 & 0.0024 \\
SN 2019bsw / ZTF19aajwkbb & 10:05:06.11 & -16:24:21.32 & WISEA J100506.20-162425.1 & 0.029 \\
SN 2019zfc / ZTF20aafezcz & 03:46:55.63 & +00:02:27.20 & SDSS J034655.55+000225.9 & 0.031 \\
SN 2020oem / ZTF20abjwntg & 15:27:29.81 & +03:46:44.81 & WISEA J152729.72+034646.8 & 0.042 \\
SN 2020abah / ZTF20actkutp & 11:36:53.47 & +21:00:01.08 & CGCG 127-002 & 0.030 \\
SN 2021mju / ZTF21aasksnl & 16:41:47.59 & +19:21:53.57 & WISEA J164148.29+192203.6 & 0.028 \\
SN 2021skm / ZTF21abjcliz & 16:16:56.05 & +21:48:35.73 & 2MASX J16165615+2148359 & 0.031 \\
SN 2021wun / ZTF21abtephz & 15:46:31.96 & +25:25:44.50 & WISEA J154631.96+252545.7 & 0.023\\
\enddata
\tablecomments{Selection criteria detailed in \S\ref{sec:sample selection}. \tablenotetext{a}{obtained from NED}}
\end{deluxetable*}

Transients that pass the CLU selection process are then systematically assigned to spectroscopic follow-up for classification based on their apparent magnitude. Transients brighter than 19 mag are assigned to the robotic Spectral Energy Distribution Machine (SEDM; \citealt{Blagorodnova2018,Rigault2019}) on the Palomar 60-inch (P60) telescope, and transients between 19 and 20 mag or coincident with bright galaxy nuclei such that the low resolution of SEDM would be insufficient to separate host contamination are assigned to the Double Beam Spectrograph (DBSP; \citealt{DBSP}) on the 200-inch Hale telescope (P200) at Palomar Observatory. If it was not possible to classify a transient using DBSP, usually due to weather or significant host contamination in faint sources, the Low-Resolution Imaging Spectrograph (LRIS; \citealt{LRIS}) on the Keck-I telescope is used as well.

\subsection{The Sample of Long-Rising Type II Supernovae}
\label{subsec:SN II sample}

We searched the sample of all 3444 transients saved to the CLU experiment on or prior to 2021 December 31 for long-rising SNe II using a tiered approach. For the sample selection process, we followed the \citet{Taddia2016} definition of ``long-rising": the $r$-band light curve rising for $>$40 days before peak. 

We first selected all 1026 spectroscopically classified SNe II and downloaded 5$\sigma$ alert photometry from the Fritz portal. We applied a set of filters to this 5$\sigma$ photometry to select those that may be long-rising and require further inspection, grouping the SNe into ``gold," ``silver," and ``bronze" categories based on decreasing filter strictness. Each SN was considered first in only $r_{ZTF}$ alert photometry and reconsidered in $g_{ZTF}$ only if it did not pass any of the criteria. 

For the ``gold" category, we required that the time between the peak detection and first detection $\Delta t_{det}$ be between 40-150 days and that the magnitude difference between the first detection and peak detection $\Delta m_{det}$ be $>$1.8 mag. A SN was also placed into the ``gold" category if $\Delta m_{det}$$>$1 mag, it met the 150d $>\Delta t_{det}>$ 40d criterion, there were at least 3 detections prior to peak, and it had an upper limit $<$15 days before the first detection. SNe with only 150d $>\Delta t_{det}>$ 40d and $\Delta m_{det}$$>$1 mag were placed into the ``silver" category. Finally, the ``bronze" category included SNe with either: 150d $>\Delta t_{det}>$ 40d but $\Delta m_{det}$$<$1 mag, at least 3 detections prior to peak but $>$40 d between the last upper limit and first detection, or $-0.015 \leq \nicefrac{\Delta m_{40d}}{\Delta t_{40d}} \leq -0.025$ (where $\nicefrac{\Delta m_{40d}}{\Delta t_{40d}}$ is a rough estimate of the light curve slope during the last $\approx$40 d prior to peak). The last ``bronze" criterion aimed to select SNe whose long rises were not detected in 5$\sigma$ photometry but might be detected in 3$\sigma$ photometry. In total, this step yielded 264 candidate SNe (41 ``gold," 5 ``silver," and 218 ``bronze"), $\approx$25\% of the total CLU SN II sample.

Next, we inspected the spectroscopy of the 264 remaining SNe II and removed 38 known SNe IIn and IIb from the list of candidates. We also removed ZTF20abefdgo from the sample at this stage because the source had $z>0.05$ based on its H$\alpha$ feature, placing it outside the ZTF-I CLU volume (see \S\ref{sec:ZTF CLU}). It most likely occurred in a background galaxy within 100$\arcsec$ of the nearby one, leading to an incorrect cross-match. We requested 3$\sigma$ forced photometry at the transient location through the ZTF forced photometry service \citep{Masci2018} of the remaining 225 candidates. For the final selection, the forced photometry light curves were passed through a second filter with tighter constraints. The forced photometry selection criteria at this stage were:
\begin{enumerate}
    \item The time between the first detection and peak in $r$-band photometry is greater than 40 days, as described in \citealt{Taddia2016}). The peak must also be at least 1 magnitude brighter than the first detection and there must be a pre-explosion upper limit 15 days before the first detection. In cases where the observed rise is more than 1.8 magnitudes, the upper limit requirement is waived. This is our basic definition for a well-defined long rise.
    \item The change in magnitude over the second half of the rise is at least 10\% of the total change in magnitude during the rise. The first detection is also within 200 days of the peak. Combined with the magnitude difference requirement in (1), these requirements filter out the majority of SN IIP plateaus.
    \item The second detection must be within 40 days of the first detection. This requirement helps remove most scattered spurious detections prior to the transient event.
\end{enumerate}
\begin{deluxetable*}{ccccc}[!ht]
\tablewidth{0pt}
\tablecaption{Discovery Data of CLU Long-rising SNe II \label{table: discovery info}}
\tablehead{
\colhead{} & \colhead{Last Nondetection}\vspace{-0.4cm} & \colhead{} & \colhead{Discovery Date} & \colhead{} \\
\colhead{SN Name} & \colhead{}\vspace{-0.4cm} & \colhead{Limiting Mag} & \colhead{} & \colhead{Discovery Mag}\\
\colhead{} & \colhead{(MJD)} & \colhead{} & (MJD) & \colhead{}
}
\startdata
SN 2018cub & 58214.82 & 22.46 & 58217.83 & 20.85 \\
SN 2018lrq & 58218.76 & 22.37 & 58221.76 & 21.86 \\
SN 2018ego & 58246.88 & 21.43 & 58249.78 & 20.63 \\
SN 2018lsg & ... & ... & 58312.90 & 20.89 \\
SN 2018imj & 58375.02 & 22.68 & 58385.02 & 19.59 \\
SN 2018hna & 58353.64 & 22.28 & 58423.02 & 15.83 \\
SN 2019bsw & 58472.94 & 24.09 & 58492.44 & 21.11 \\
SN 2019zfc & 58804.83 & 24.77 & 58833.77 & 21.36 \\
SN 2020oem & 58979.82 & 21.96 & 58985.77 & 21.01 \\
SN 2020abah & 59174.02 & 21.81 & 59177.05 & 20.26 \\
SN 2021mju & 59283.01 & 22.95 & 59291.96 & 21.33 \\
SN 2021skm & 59387.90 & 22.67 & 59390.79 & 20.37 \\
SN 2021wun & 59423.73 & 23.36 & 59426.75 & 20.58 \\
\enddata
\tablecomments{All magnitudes in $r_{ZTF}$. The data in this table are derived from $3\sigma$ ZTF force photometry at the location of each transient (see Table \ref{table: SN info}). The ZTF difference imaging pipeline triggers alerts only on $5\sigma$ detections, so reported values in this table generally occur prior to the first alert for each source.}
\end{deluxetable*}

\begin{figure*}[!ht]
    \centering
    \includegraphics[scale=0.75]{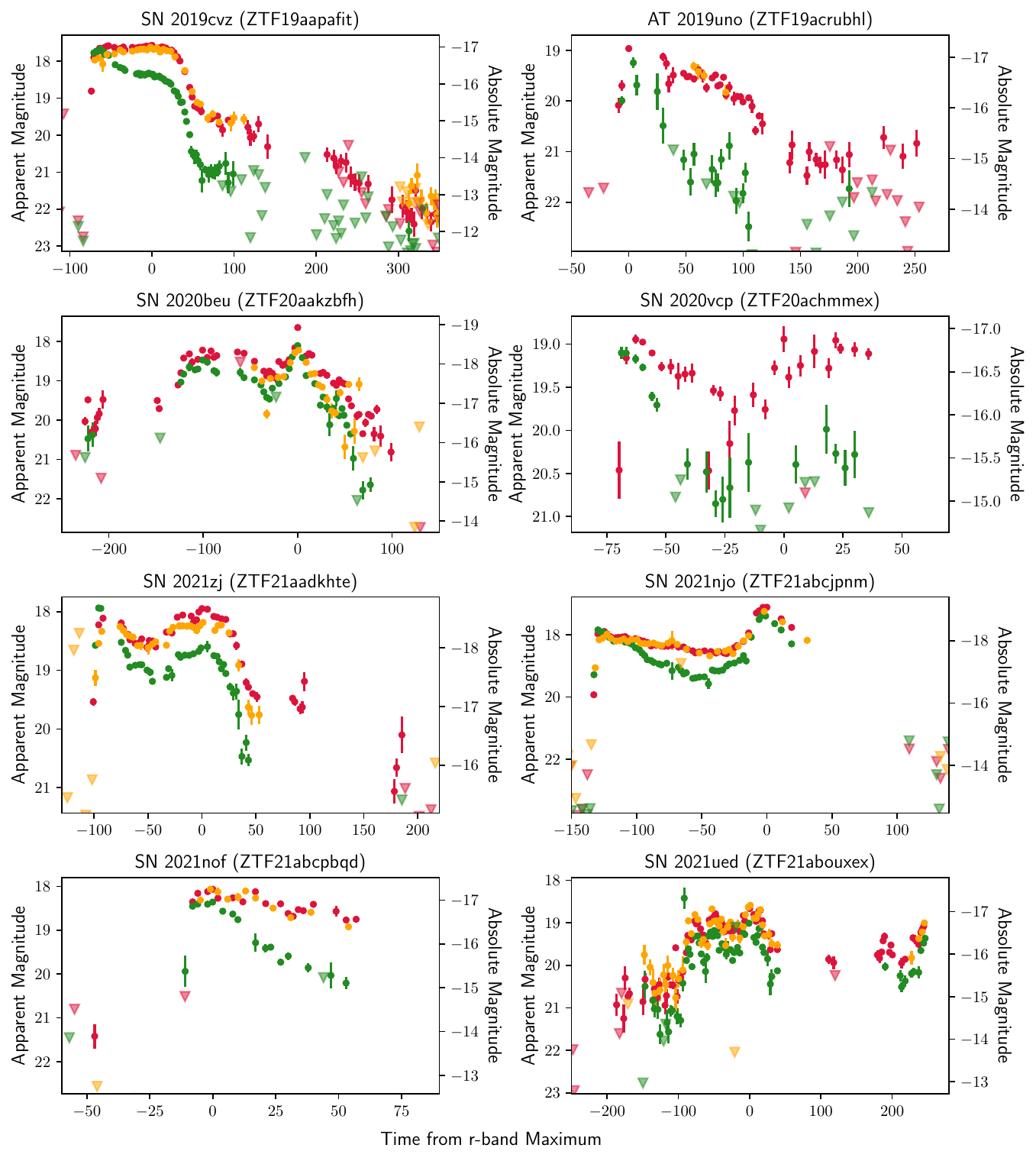}
    \caption{Optical light curves of eight candidate SNe rejected after visual inspection in 3$\sigma$ ZTF $gri$ forced photometry. The red, green, and orange points denote the $r$-, $g$-, and $i$-bands respectively. Triangles indicate 3$\sigma$ upper limits. Inverse-variance weighted binning was applied to the photometry using a bin size of 2 d. Times are reported in the observer frame.}
    \label{fig: reject LCs}
\end{figure*}

The light curves of the 20 candidates passing this second photometry filter were then plotted and inspected by eye. Each light curve was also fit to a SN 1987A model using the \texttt{sncosmo} Python package (see \S\ref{sec:photometry} for more details) to aid in visually identifying the characteristic long-rising, dome-shaped light curve. Twelve candidates remained after visual inspection of their light curves. Eight candidates were rejected because they did not show the characteristic dome-shaped light curve. We plot these light curves in Figure \ref{fig: reject LCs}. SN 2019cvz (ZTF19aapafit) and AT 2019uno (ZTF19acrubhl) appear to be normal SN IIP. SN 2021nof (ZTF21abcpbqd) has an early detection that may indicate precursor activity or may be spurious, as the SN is located near the bright center of its host, but regardless does not show a 1987A-like light curve shape. The double-peaked behavior of the SN 2020vcp (ZTF20achmmex), SN 2021zj (ZTF21aadkhte), and SN 201njo (ZTF21abcjpnm) light curves is inconsistent with an early bright bump seen in some 87A-like SNe from the literature such as SN 2004ek \citep{Taddia2016}; these SNe may be members of a different class of SNe II with unusually wiggly plateaus. The extended activity and central locations of SN 2020beu (ZTF20aakzbfh) and SN 2021ued (ZTF21abouxex) in their host galaxies suggest that they may actually be AGN.

We also recovered one additional long-rising SN II whose 5$\sigma$ alert photometry was too sparse to pass our earlier selection criteria. Tzanidakis et al. (in prep) identified 13 SNe II in CLU with substantially longer rise times than the rest of their sample based on a fit to a parametric light curve model (see \citealt{Villar2019}, Eq. 1). We obtained forced photometry light curves for these SNe II for manual inspection and found that the light curve of SN 2018ego (ZTF18ablhrpz) passes the selection criteria described previously. 

The defining characteristic of 1987A-like SNe in the literature is their long dome-shaped rises, so photometry was used as the primary method for sample selection. The spectra were used to confirm the presence of broad H$\alpha$ and thus the SN II classification. 

Our final sample consists of 13 long-rising SNe II with well-defined light curves identified from the ZTF CLU experiment between 2018 June 1 and 2021 December 31 (Table \ref{table: SN info}). The discovery data are presented in Table \ref{table: discovery info}. We present the full optical ZTF forced photometry light curves and discuss the photometric properties of the sample in detail in \S\ref{sec:photometry}. The available spectra of the SNe were checked for a reasonably secure SN II classification, and the spectroscopic properties of the sample are further discussed in \S\ref{sec:spectroscopy}.

\begin{deluxetable*}{ccccc}[!th]
\tablewidth{0pt}
\tablecaption{Rise Times and Peak magnitudes of CLU Long-rising SNe II \label{table: rise times and max mags}}
\tablehead{
\colhead{} & \colhead{$t_{rise}$}\vspace{-0.4cm} & \colhead{$m_{max,g}$} & \colhead{$m_{max,r}$}& \colhead{$m_{max,i}$} \\
\colhead{SN Name} & \colhead{}\vspace{-0.4cm} & \colhead{} & \colhead{} & \colhead{} \\
\colhead{} & \colhead{(d)} & \colhead{(mag)} & \colhead{(mag)} & \colhead{(mag)}
}
\startdata
SN 2018cub & 86 & 20.11 & 19.28 & -- \\
SN 2018lrq & 88 & 20.94 & 19.46 & 19.24 \\
SN 2018ego & 72 & 18.93 & 18.67 & 19.40 \\
SN 2018lsg & 76\tablenotemark{a} & 16.61 & 18.93 & 19.24 \\
SN 2018imj & 74 & 18.95 & 17.78 & -- \\
SN 2018hna & 80\tablenotemark{a,b} & 14.59 & 13.99 & 15.05 \\
SN 2019bsw & 87 & 19.63 & 18.74 & -- \\
SN 2019zfc & 81\tablenotemark{b} & 19.89 & 19.09 & -- \\
SN 2020oem & 96 & 19.96 & 19.04 & -- \\
SN 2020abah & 84 & 19.98 & 19.13 & -- \\
SN 2021mju & 78 & 20.72 & 19.97 & 20.01 \\
SN 2021skm & 68 & 20.35 & 18.77 & 18.84 \\
SN 2021wun & 67 & 19.39 & 18.44 & 18.62 \\
\enddata
\tablecomments{$t_{rise}$ is defined as the time from the midpoint of the last non-detection and first detection in the $r$-band to the peak time of the fitted \texttt{sncosmo} SN 1987A model (unless otherwise noted). Values calculated using binned ZTF forced photometry with 2 d bins, except for SNe 2018cub and 2018lrq (3 d bins) and SN 2018lsg (4 d bin).
\tablenotetext{a}{Did not have an upper limit close to first detection, so only the time of first detection was used.} 
\tablenotetext{b}{The empirical $r$-band peak time was used instead, because in some instances the model peak time would be shifted relative to the data peak due to small differences in light curve shape.}}
\end{deluxetable*}

\section{Photometry}
\label{sec:photometry}

\begin{figure*}[!hp]
    \centering
    \includegraphics[scale=0.75]{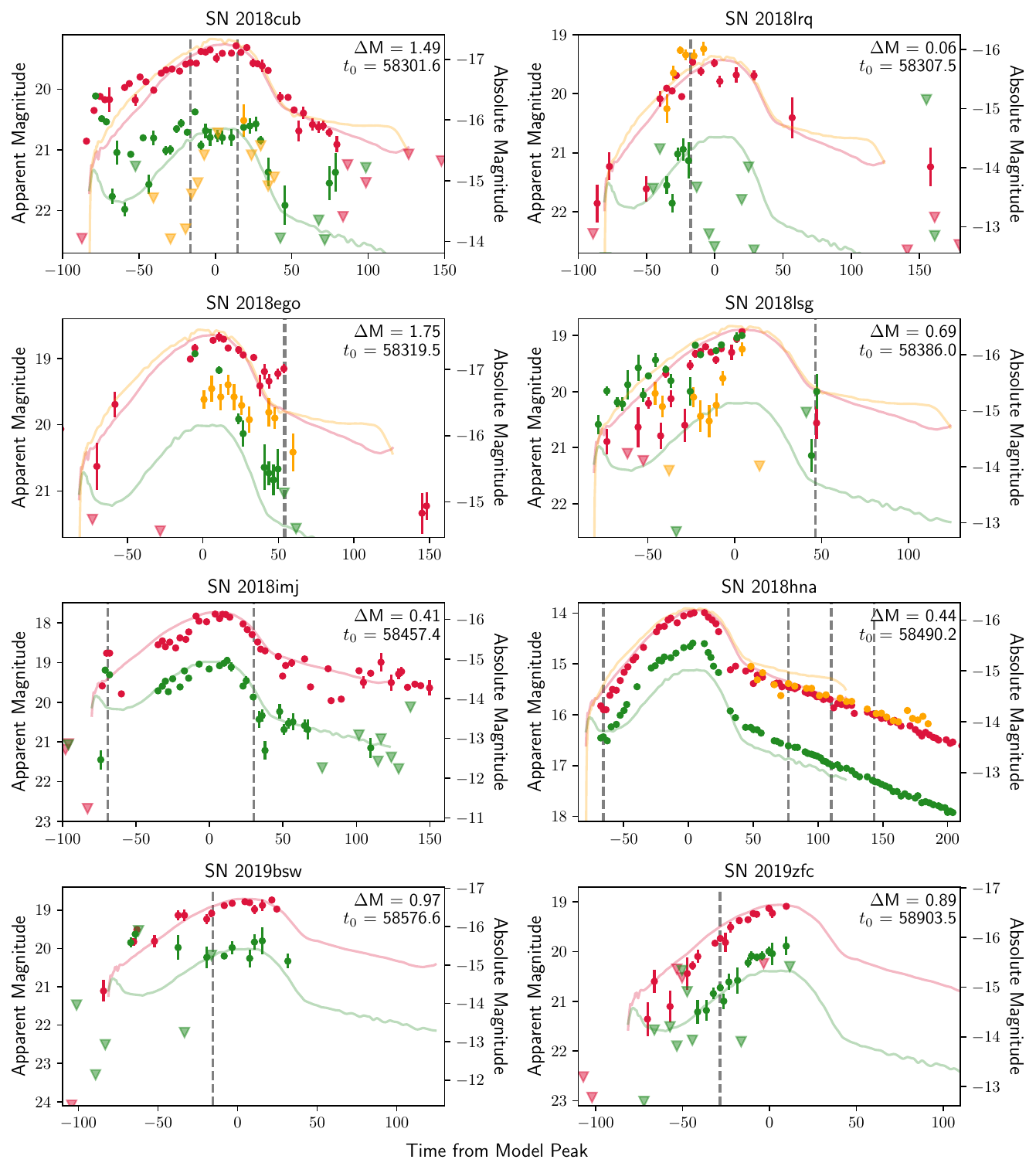}
    \caption{Optical light curves of the sample of long-rising SNe II in 3$\sigma$ ZTF $gri$ forced photometry. The red, green, and orange points denote the $r$-, $g$-, and $i$-bands respectively. Triangles indicate 3$\sigma$ upper limits. The solid lines are synthetic ZTF photometry generated from the \texttt{sncosmo} SN 1987A model scaled to the best-fit magnitude and phase for each observed light curve. Following the convention of \texttt{sncosmo} models, times are reported in the observer frame. The dashed gray lines indicate spectroscopic epochs. For SN 2018hna, only epochs of optical spectra taken by ZTF are shown. Inverse-variance weighted binning was applied to the photometry using a bin size of 2 d except for SNe 2018cub and 2018lrq (3 d), and SN 2018lsg (4 d). On the upper right of each plot, we report $\Delta$M and $t_0$. $\Delta$M is the magnitude difference between the best-fit peak magnitude and the peak ZTF $r$-band magnitude of SN 1987A generated from synthetic \texttt{sncosmo} photometry, with positive values being brighter than SN 1987A. $t_0$ is the best-fit model peak time in MJD.
    The binned photometry and upper limits used to generate this figure will be available online. 
    Additional optical light curves of the sample are in Figure \ref{figures: gold sample full LCs 2}.}
    \label{figures: gold sample full LCs 1}
\end{figure*}

\begin{figure*}[!ht]
    \centering
    \includegraphics[scale=0.75]{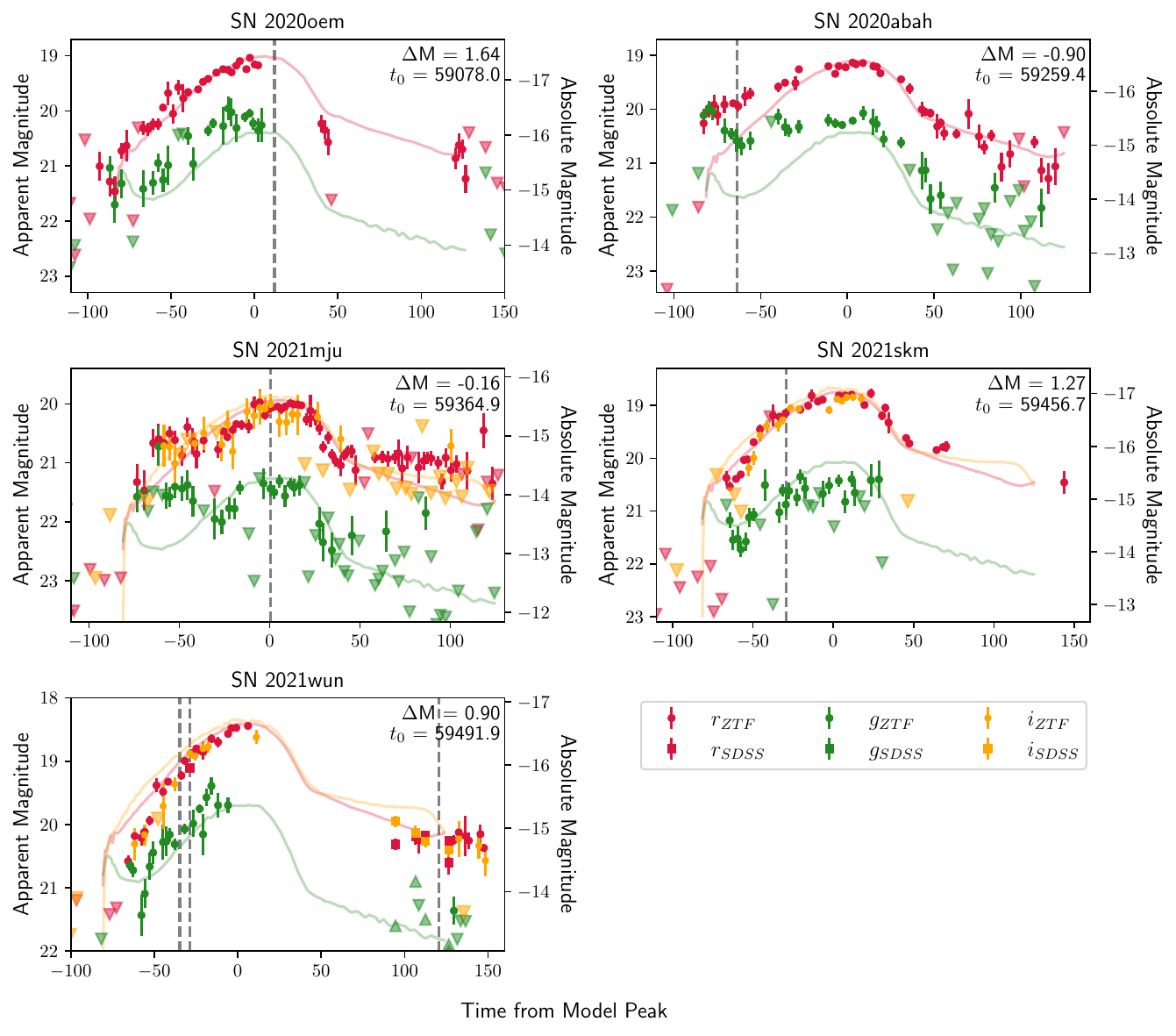}
    \caption{Optical light curves of the sample of long-rising SNe II in 3$\sigma$ ZTF $gri$ forced photometry. Times reported in the observer frame. Inverse-variance weighted binning using a bin size of 2 d was applied for all ZTF photometry in this plot. Symbols and line styles are the same as in Figure \ref{figures: gold sample full LCs 1}. Additional photometry in the Sloan Digital Sky Survey (SDSS) passbands was obtained with P60 for SN 2021wun and is indicated by square points. 
    The binned photometry and upper limits used to generate this figure will be available online.
    The figures for the rest of the sample are also shown in Figure \ref{figures: gold sample full LCs 1}.}
    \label{figures: gold sample full LCs 2}
\end{figure*}

\begin{figure*}[!t]
    \centering
    \includegraphics[scale=0.5]{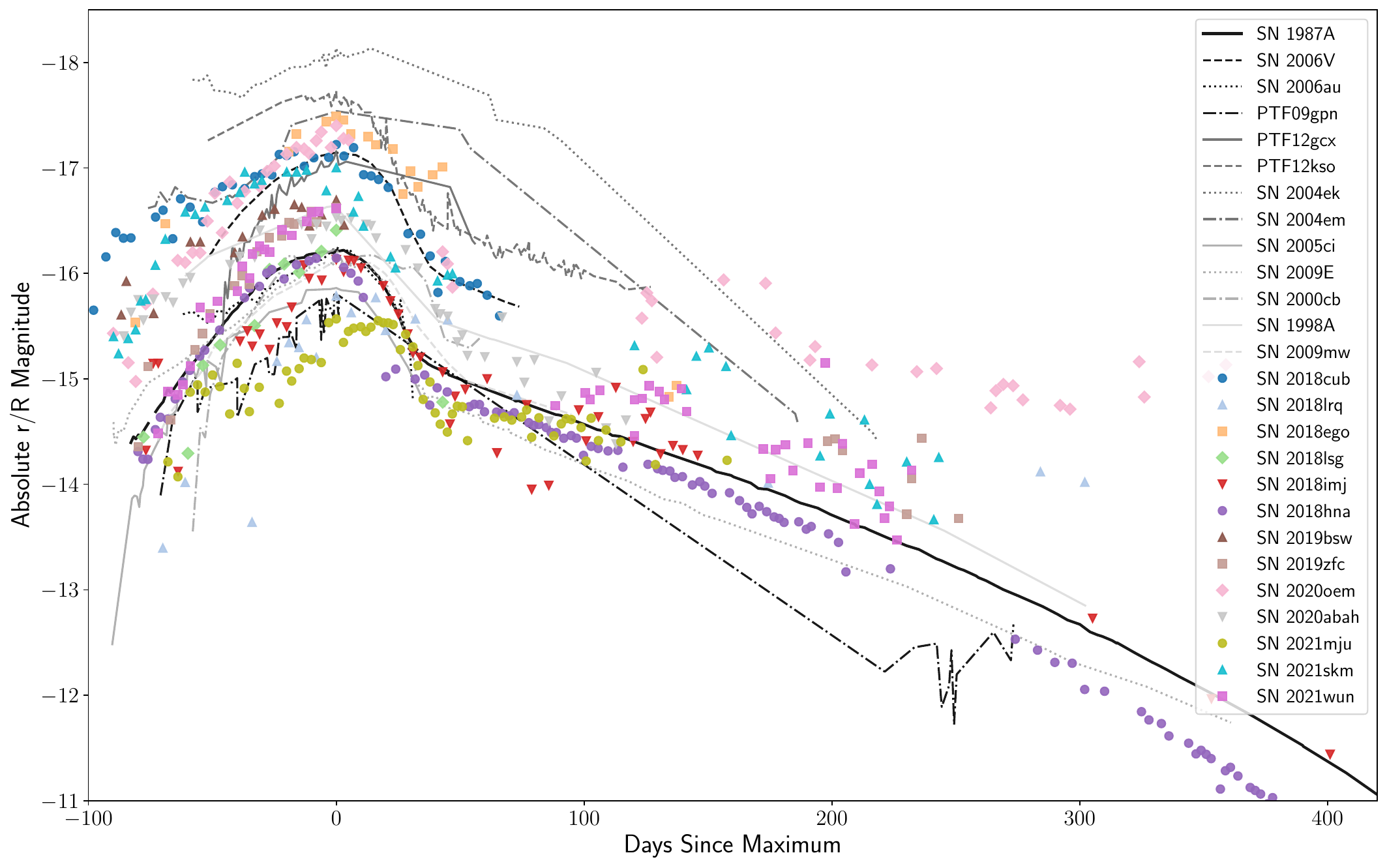}
    \caption{$r/R$-band absolute magnitude light curves of the CLU sample, plotted alongside SN 1987A and various other long-rising SNe II from the literature. The available photometry is most complete in the ZTF $r$-band for the ZTF sources and $r/R$/unfiltered bands for the literature sources. The data from SN 1987A here is $R$-Cousins photometry \citep{Hamuy1990} publicly available from WISeREP \citep{wiserep}, rather than the \texttt{sncosmo} synthetic photometry used elsewhere, for the extended photometric coverage at $\gtrapprox$130 d after peak. The photometry for the literature sources are from \citealt{Taddia2012} (SNe 2006V, 2006au), \citealt{Taddia2016} (PTF sources, SNe 2004ek, 2004em, 2005ci), \citealt{Pastorello2012} (SN 2009E), \citealt{Kleiser2011} (SN 2000cb), \citealt{Pastorello2005} (SN 1998A), and \citealt{Takats2016} (SN 2009mw).
    The ZTF photometry and upper limits of the CLU sample will be available online.}
    \label{fig:gold r-band LCs with literature}
\end{figure*}

All long-rising SNe II in our sample were observed primarily in the ZTF $r$ and $g$ filters, with some $i$-band detections for most sources as well. Light curves from ZTF forced photometry at the target location showing the long rise and characteristic 1987A-like dome shape are plotted in Figures \ref{figures: gold sample full LCs 1} and \ref{figures: gold sample full LCs 2}. Additional late-time photometry ($\gtrapprox$150 d after peak) is shown in Figure \ref{fig:gold r-band LCs with literature}. Phases are reported in the observer frame.

Data for SN 1987A were obtained with the built-in template \texttt{v19-1987a} \citep{Vincenzi2019} in the \texttt{sncosmo} Python package \citep{sncosmo}. The \texttt{sncosmo} package also provides functions for light-curve fitting and generating synthetic photometry from a model in a given bandpass. For each SN in the sample, the SN 1987A model was fit to the observed light curve; the fitted model light curves are plotted as solid lines in Figures \ref{figures: gold sample full LCs 1} and \ref{figures: gold sample full LCs 2}. The maximum apparent magnitude in each ZTF bandpass and $r$-band rise times of the sample are reported in Table \ref{table: rise times and max mags}.

In this section, all dates and times are reported in UT. Absolute magnitudes are calculated from the luminosity distance at the redshift of each SN, assuming $H_0=67.66$ km s$^{-1}$ Mpc$^{-1}$ \citep{cosmologyparams}.

\subsection{The CLU Sample of Long-rising SNe II}
\subsubsection{SN 2018cub / ZTF18aaikcbb}
SN 2018cub was first detected in the ZTF difference imaging pipeline (ie. the first alert packet) on 2018 April 10 08:23:02 at J2000 coordinates $\alpha=$ 15:05:42.45, $\delta=+$60:47:43.85 with magnitude $r_{ZTF}= 20.65 \pm 0.15$. The source was first reported to the Transient Name Server (TNS) by ATLAS on 2018 June 23 \citep{2018cub_TNS}, although we later reported our earlier April 2018 ZTF detection on 2020 April 15. Based on ZTF forced photometry, the first 3$\sigma$ detection was on 2018 April 9 (MJD = 58217.83) at a magnitude of $r_{ZTF}=20.85$ mag. The SN shows an early shock cooling feature that is brighter than that of SN 1987A (scaled to the same peak magnitude), leading to a shallower rise during the first $\approx$40 days. After $\approx$40 days, the $r$-band light curve closely follows the expected 1987A-like dome shape. The $g$-band photometry is fairly scattered. SN 2018cub is located 8\farcs3 from the center of its host galaxy WISEA J150541.98+604751.4 at a redshift of $z=0.044$, corresponding to a physical separation of $\approx$8 kpc. This makes it the most distant long-rising SN II in our sample; within our sample limit of $z=0.05$, only PTF12gcx, at $z=0.045$ \citep{Taddia2016}, is more distant.

\subsubsection{SN 2018lrq / ZTF18aaxzlmy}
SN 2018lrq was first detected in the ZTF difference imaging pipeline on 2018 June 5 04:49:47 at J2000 coordinates $\alpha=$ 13:34:51.40, $\delta=+$34:03:20.46 with magnitude $r_{ZTF}=19.83 \pm 0.17$. The transient was reported to TNS on 2020 April 15 \citep{2018lrq_2018slg_TNS}. ZTF forced photometry at this location shows a much earlier $3\sigma$ detection on 2018 April 13 (MJD = 58221.76) at $r_{ZTF}=21.86$ mag. Early detections are unusually sparse given ZTF's high cadence, but still generally show a long rise similar in shape to that of SN 1987A in the $r$-band. There are very few detections even in forced photometry during the fade, but the only two 3$\sigma$ $r$-band detections suggest that the transient faded at a similar rate to SN 1987A. SN 2018lrq is located 2\farcs0 from the center of the galaxy UGC 08561 at $z=0.025$, corresponding to a physical separation of $\approx$1 kpc.

\subsubsection{SN 2018ego / ZTF18ablhrpz}
SN 2018ego was first detected in the ZTF difference imaging pipeline on 2018 August 3 03:56:45.89 at J2000 coordinates $\alpha=$ 15:52:54.02, $\delta=+$19:58:05.33 with magnitude $r_{ZTF}=18.77 \pm 0.11$. The source was reported to TNS earlier, on 2018 July 25, with magnitude $i_{SDSS}=18.82\pm0.06$ by Pan-STARRS \citep{2018ego_TNS}. ZTF forced photometry at the transient location reveals an early $3\sigma$ detection on 2018 May 11 (MJD = 58249.78) at $r_{ZTF}=20.63$ mag. There are few detections prior to peak for this transient, but the $3\sigma$ detections in forced photometry are generally consistent with a long rise to peak in the $r$-band. The $i$-band light curve, although only detected in $3\sigma$ forced photometry near peak, appears to peak at a similar time to the $r$-band, and is consistently $\sim$1 mag fainter than $r$ detections at similar phase. The $g$-band light curve, on the other hand, appears to have peaked about 2 weeks prior to the $r$- and $i$-band light curves, although there were no detections in $g$ during the rise in this filter. There is an unusual rise in $r_{ZTF}$ and $i_{ZTF}$ (more notable in the $r$-band) about a month after peak; however, the transient then went behind the sun and there are no further detections until the transient was recovered briefly near the end of its fade. The SN is located in the far outskirts (26\farcs3 of separation) of the galaxy 2MASX J15525218+1958107, which has a redshift of $z=0.037$ (physical separation $\approx$21 kpc).

\subsubsection{SN 2018lsg / ZTF18abtjmns}
SN 2018lsg was first detected in the ZTF difference imaging pipeline on 2018 September 5 06:29:17 at J2000 coordinates $\alpha=$ 20:46:44.59, $\delta=-$ 01:22:07.66 with magnitude $g_{ZTF}= 19.29 \pm 0.09$. The transient was reported to TNS on 2020 April 15 \citep{2018lrq_2018slg_TNS}. In 3$\sigma$ ZTF forced photometry, the first detection was 2018 July 13 (MJD = 58312.90) with magnitude $g_{ZTF}=20.89$ mag. The earliest photometry is fairly scattered but overall a long rise is visible in both the $g$- and $r$-bands. This SN is very blue compared to SN 1987A, so its $g$-band photometry matches the \texttt{sncosmo} SN 1987A $r$-band photometry better in shape. However, there is a slight dip in the rise in the $g$- and $r$-bands around $-$30 days, followed in the $i$-band around $-$20 days. There is only one detection after peak, after the SN had already faded significantly, so the second part of the expected dome shape of the light curve was not well documented. SN 2018lsg is located only 0\farcs1 from the center of its host galaxy CGCG 374-012 at $z=0.025$, corresponding to a physical separation of $\approx$0.04 kpc.

\subsubsection{SN 2018imj / ZTF18abytyif}
SN 2018imj was first detected in the ZTF difference imaging pipeline on 2018 September 23 12:30:00.58 at J2000 coordinates $\alpha=$ 06:51:05.75, $\delta=+$12:54:55.78 with magnitude $r_{ZTF}=20.14 \pm 0.22$. It was reported to TNS on 2018 November 15 by the Lick Observatory Supernova Search \citep{2018imj_TNS}. The first $r$-band detection in $3\sigma$ ZTF forced photometry is on the same date, although the forced photometry detection has a magnitude of $r_{ZTF}=19.59$. There is also a slightly earlier $g$-band detection in forced photometry on 2018 September 22 (MJD = 58383.50) at $g_{ZTF}=21.45 \pm 0.24$. The rise of SN 2018imj is well sampled starting $\approx$40 days prior to peak. The shape of the light curve of this transient is very similar to that of SN 1987A in both the rise and decline, with a slightly steeper rise just before peak. SN 2018imj is located in the outskirts of IC 0454, a SBab galaxy at redshift $z=0.013$, 24\farcs9 from the nucleus, $\approx$7 kpc. About 150 days after peak, the transient went behind the sun, and it was recovered at the end of its fade $\approx$150 days after it disappeared, about 300 days after peak. These late time detections are also generally consistent with the absolute magnitude of SN 1987A in Cousins-R photometry at late times (from WISeREP, \citealt{wiserep}).

\subsubsection{SN 2018hna / ZTF18acbwaxk}
SN 2018hna was first detected in the ZTF difference imaging pipeline on 2018 October 31 11:42:05.184 at J2000 coordinates $\alpha=$ 12:26:12.08, $\delta=+$58:18:50.77 with magnitude $r_{ZTF}= 16.41 \pm 0.03$. The SN was first discovered by Koichi Itagaki on 2018 October 22 19:33:07 and reported to TNS within two hours \citep{2018hna_TNS}. The $r$-band rise is slightly steeper at the earliest times, but otherwise the shape of the $r$-band rise mirrors that of SN 1987A. The shock cooling emission feature mentioned in \citet{Singh2019} was not observed by ZTF. SN 2018hna occurred 43\farcs5 from its host galaxy UGC 07534 at $z=0.0024$, corresponding to a physical separation of $\approx$2 kpc. Additional photometry and more detailed analysis of this SN can be found in \citet{Singh2019} and \citet{Xiang2023}.

\subsubsection{SN 2019bsw / ZTF19aajwkbb}
SN 2019bsw was first detected in the ZTF difference imaging pipeline on 2019 February 10 10:36:26 at J2000 coordinates $\alpha=$ 10:05:06.10, $\delta=-$ 16:24:21.49 with magnitude $r_{ZTF}= 19.80 \pm 0.20$. The transient was first reported to TNS by ATLAS on 2019 March 14 \citep{2019bsw_TNS}. The first detection in 3$\sigma$ forced photometry occurred about a month and a half before the TNS report on 2019 January 9 (MJD = 58492.44) with a magnitude of $r_{ZTF}=21.11$. Around peak, both the $g$- and $r$-bands follow the expected dome-like light curve shape well, and the first $r$-band detection in forced photometry is consistent with a slightly longer rise than that of SN 1987A, However, the early part of the light curve is somewhat different: photometry appearing to resemble shock cooling occurs $\approx$60 days prior to peak, which is later than this feature appears in the other long-rising SNe II in our sample. SN 2019bsw is located 4\farcs7 from the center of its host galaxy WISEA J100506.20-162425.1 at a redshift of $=0.029$, corresponding to $\approx$3 kpc.

\subsubsection{SN 2019zfc / ZTF20aafezcz}
SN 2019zfc was first detected in the ZTF difference imaging pipeline on 2020 January 14 04:43:31 at J2000 coordinates $\alpha=$ 03:46:55.64, $\delta=+$ 00:02:27.13 with magnitude $r_{ZTF}= 20.15 \pm 0.19$. The transient was first reported to TNS by Pan-STARRS on 2019 December 30 \citep{2019zfc_TNS}. We find an earlier detection in 3$\sigma$ ZTF forced photometry on 2019 December 16 (MJD = 58833.77) at a magnitude of $r_{ZTF}=21.36$. Its rise in both the $g$- and $r$-bands is slightly shorter and steeper than that of SN 1987A. The transient went behind the sun around peak so the fade of its light curve was not documented, but a few faint ($m_{r,ZTF}>21$ mag) detections were recovered in forced photometry $\approx$200 days after the apparent peak detection when it reappeared. ZTF20aafezez is 1\farcs6 from the center of its host galaxy SDSS J034655.55+000225.9 with $z=0.031$, corresponding to $\approx$1 kpc.

\subsubsection{SN 2020oem / ZTF20abjwntg}
SN 2020oem was first detected in the ZTF difference imaging pipeline on 2020 July 8 05:06:16 at J2000 coordinates $\alpha=$ 15:27:29.81, $\delta=+$ 03:46:44.93 with magnitude $r_{ZTF}= 19.71 \pm 0.18$. The transient was reported to TNS the same day \citep{2020oem_TNS}. The earliest detection in 3$\sigma$ ZTF forced photometry was on 2020 May 16 (MJD = 58985.77) and had a $r$-band magnitude of 21.01. The earliest forced photometry detections for this SN II appear to be at the end of the shock cooling phase, which occurs at a slightly earlier time compared to peak than SN 1987A (based on the \texttt{sncosmo} model) and most of the other long-rising SNe II in our sample showing the shock cooling feature in their early-time photometry. The $r$-band light curve follows the dome shape of SN 1987A well throughout the rise. Detections are sparse after peak, but they suggest that the end of dome-shaped part of the light curve occurs at a lower luminosity than expected from the SN 1987A model, while the slope of the plateau is slightly shallower. SN 2020oem is located 2\farcs0 from the center of its host galaxy WISEA J152729.72+034646.8, which has $z=0.042$ (physical separation $\approx$2 kpc).

\subsubsection{SN 2020abah / ZTF20actkutp}
SN 2020abah was first detected in the ZTF difference imaging pipeline on 2020 November 23 12:25:00 at J2000 coordinates $\alpha=$ 11:36:53.45, $\delta=+$ 21:00:01.07 with magnitude $g_{ZTF}= 20.06 \pm 0.22$. We reported this transient to TNS on 2020 November 26 \citep{2020abah_TNS}. In 3$\sigma$ ZTF forced photometry, the first detection was at 2020 November 23 (MJD = 59176.52), the same time as the pipeline alert, and had a $g$-band magnitude of 20.11. A $r$-band detection at 20.26 mag occurred within an hour of the $g$-band detection in forced photometry. Similarly to SN 2018cub, SN 2020abah has a slower rise than SN 1987A did in the first $\approx$40 days before following the domed light curve shape well afterward, through to the plateau phase. The $g$-band photometry is fairly flat after the shock cooling feature fades until after peak. SN 2020abah is located 14\farcs7 from the center of CGCG 127-002, with a redshift of $z=0.030$, which corresponds to a physical separation of $\approx$10 kpc.

\subsubsection{SN 2021mju / ZTF21aasksnl}
SN 2021mju was first detected in the ZTF difference imaging pipeline on 2021 April 4 10:56:49 at J2000 coordinates $\alpha=$ 16:41:47.59, $\delta=+$ 19:21:53.35 with magnitude $r_{ZTF}= 20.73 \pm 0.20$. The transient was reported to TNS on 2021 May 15 \citep{2021mju_TNS}. The first detection in 3$\sigma$ ZTF forced photometry is at 2021 March 18 (MJD = 59291.96) with $r_{ZTF}=21.33$ mag. The shape of the $r$-band light curve generally follows the expected dome shape well, except for what is likely the shock cooling bump prior to $\approx$40 days before peak. This feature occurs somewhat late in the early-time light curve, similarly to SN 2019bsw. The rate at which this SN fades in the plateau phase is also noticeably slower than that of SN 1987A. The $g$-band photometry shows an interesting rise during the plateau phase as well; however, the detections are faint, with dispersed upper limits in the forced photometry, so we caution that the $g$-band detections showing this trend may not be particularly reliable. SN 2021mju is located 13\farcs9 from the center of its host WISEA J164148.29+192203.6, which has a redshift of $z=0.028$ (physical separation $\approx$9 kpc).

\subsubsection{SN 2021skm / ZTF21abjcliz}
SN 2021skm was first detected in the ZTF difference imaging pipeline on 2021 July 4 05:29:00 at J2000 coordinates $\alpha=$ 16:16:56.06, $\delta=+$ 21:48:35.65 with magnitude $r_{ZTF}= 20.25 \pm 0.21$. The transient was reported to TNS on 2021 July 7 \citep{2021skm_TNS}. The first 3$\sigma$ ZTF forced photometry detection was on 2021 June 25.8 (MJD = 59390.79) with $r_{ZTF} = 20.37$ mag. The $r$-band rise in the first $\approx$30 days is faster than that of SN 1987A, but by $\approx$40 days prior to peak the light curve closely follows the expected dome shape to the start of the plateau. Signs of a shock cooling feature can also be seen at very early times. At the plateau phase of the light curve the $r$-band appears to begin to rise slightly, but the next detection at $\approx$150 days after peak suggests that the SN then faded as expected. SN 2021skm occurred 1\farcs4 from 2MASX J16165615+2148359, which has $z=0.031$, and corresponds to a physical separation of $\approx$1 kpc.

\subsubsection{SN 2021wun / ZTF21abtephz}
SN 2021wun was first detected in the ZTF imaging pipeline on 2021 August 14 05:27:43.78 at J2000 coordinates $\alpha=$ 15:46:31.96, $\delta=$ 25:25:44.50 with magnitude $r_{ZTF}=19.94\pm 0.22$. The transient was reported to TNS on 2021 August 23 \citep{2021wun_TNS}. ZTF forced photometry at this location shows an earlier 5$\sigma$ detection at 2021 July 31 (MJD = 59426.75) with a magnitude of $r_{ZTF} = 20.58$. Due to a combination of poor weather and the transient going behind the sun, the full extent of the light curve was not able to be documented, but available data shows a well-observed rise of at least 72 days. The shape of the $r$-band rising light curve closely follows that of SN1987A: it is slightly steeper for the first $\approx$30 days, but then is practically identical for the remaining $\approx$50. The $g$- and $i$-band light curves follow this shape as well. SN 2021wun is located in the galaxy SDSS J154631.94+252545.5 ($z=0.023$) at an offset of 0\farcs5, corresponding to a physical separation of $\approx$0.08 kpc.

\subsection{Color Evolution}

\begin{figure}[!t]
    \centering
    \includegraphics[scale=0.39]{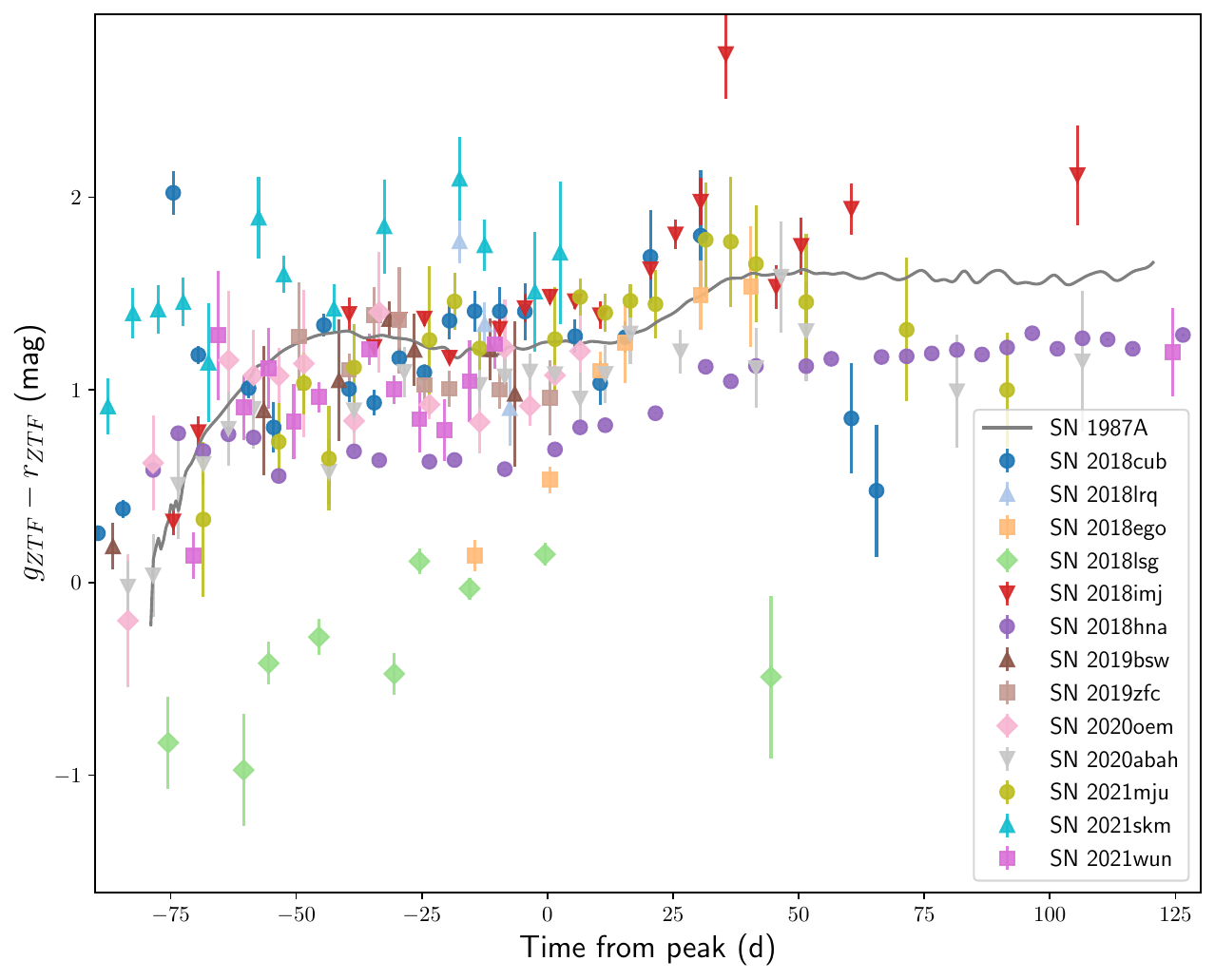}
    \caption{$g-r$ color evolution of the long-rising SNe II. Each point was calculated as $\bar{g} - \bar{r}$, where $\bar{g}$ is the inverse-variance weighted mean of all $3\sigma$ forced photometry detections in the ZTF $g$-band in a 5-day time bin, and similarly for $\bar{r}$ with ZTF $r$-band detections. The values for SN 1987A are derived from synthetic photometry in the ZTF $g$- and $r$-bands using \texttt{sncosmo}.}
    \label{fig: g-r color plot}
\end{figure}

The $g-r$ color evolution of the CLU sample is plotted in Figure \ref{fig: g-r color plot} alongside that of SN 1987A (calculated from synthetic \texttt{sncosmo} photometry). We corrected for Galactic extinction using the maps in \citet{Schlafly2011} and the extinction law of \citet{Fitzpatrick1999} assuming $R_v=3.1$ as implemented in the NED extinction calculator tool \citep{extinctioncalccite}. ZTF magnitudes were converted to PS1 magnitudes and back following the relations presented in \citet{Medford2020} to use the total absorption $A_\lambda$ in PS1 magnitudes as returned by calculator.

The color evolution of SNe 2018cub, 2018imj, 2019bsw, 2020oem, 2020abah, 2021mju, and 2021wun are all very similar to each other and that of SN 1987A. All of these sources are within $\approx$0.5 mag in color of each other at most epochs and show a fast reddening at early times ($\lessapprox$50 days before peak), a feature that is likely a sign of shock cooling. In SNe 2018cub and 2019bsw, this feature is somewhat less pronounced as these SN are already redder at $\approx$80 days before peak than SN 1987A and the other sources showing evidence of shock cooling. 

After $\approx$50 days before peak, SNe 2018cub and 2018imj show the most similar $g-r$ colors to each other and SN 1987A, although the $g$-band photometry is somewhat scattered for SN 2018cub. SN 2018imj also shows an odd reddening between $\approx20-30$ days after peak before resettling into a more expected $g-r$ color.

SNe 2019bsw, 2020oem, and 2021wun are all overall somewhat bluer than SNe 1987A, 2018cub, and 2018imj prior to peak. SN 2019zfc had no $g$-band detections early enough to show the shock cooling feature but is of a similar color. SN 2020abah shows the same color as SN 1987A during the shock cooling phase before suddenly becoming much bluer at $\approx$40 days before peak, and then reddening again to a color similar to the other SNe in our sample that are somewhat bluer than SN 1987A. 

SN 2018hna is $\approx$0.5-0.7 mag bluer than SN 1987A at all times. In ZTF photometry, we do not see evidence of shock cooling, but the feature can be seen in earlier photometric detections of this SN from other programs (see \citealp{Singh2019}). SN 2018lsg is noticeably the bluest SN in our sample; while it exhibits a gradual reddening prior to peak, it remains $\gtrapprox$1 mag bluer than SN 1987A and most of the rest of our sample at all times. On the other hand, SN 2021skm is the reddest SN in our sample, usually $\approx$0.7 mag redder than SN 1987A. It shows evidence of shock cooling at early times as well, but this feature is also redder than in the other SNe exhibiting it in our sample.

SNe 2018lrq and 2018ego show unusual color evolution compared to the rest of our sample. SN 2018lrq becomes more blue as it approaches its peak, which is different to most SNe which redden as they evolve. SN 2018ego exhibits a strong reddening beginning $\approx$15 days before peak, reaching a color similar to that of SN 1987A at the same phase $\approx$30 days after peak. However, these two SNe also have among the sparsest $g$-band photometry in our sample, so their trends in color evolution may not be very reliable.

\subsection{Comparison to Other Long-Rising SNe II}

We also compare our sample to other notable long-rising SNe II in the literature. SN 2018hna is the only previously published long-rising SN II of our ZTF CLU sample; we point to \citet{Singh2019}, \citet{Tinyanont2021}, and \citet{Xiang2023} for more complete and detailed analysis of this supernova. Our sample spans a range of absolute magnitudes in the $r$-band from $-15.6$ to $-17.5$ mag, which is broadly reflective of the range in luminosity among long-rising SNe II as a whole. 

On the bright end, SN 2018ego, which peaked at $-17.5$ mag in the $r$-band, and SN 2020oem, which peaked at $-17.4$ mag, are the most luminous SNe in our sample. Their luminosities are most comparable to PTF12kso and SN 2004em, but neither are brighter than SN 2004ek. All three of these literature SNe were published in \citet{Taddia2016} and noted to be the most luminous long-rising type II SNe found at the time. SNe 2018cub ($M=-17.2$) and 2021skm ($M=-17.0)$ round out the rest of the brightest long-rising CLU SNe II.

SNe 2018imj and 2018hna, which both peak at $-16.1$ in $r$, are overall the most similar in luminosity to SN 2009mw, SN 2009E, SN 2006au, SN 2000cb, and SN 1987A itself. SNe 2018lrq and 2021mju, peaking at $-15.8$ and $-15.6$ mag respectively in the $r$-band, are two of the faintest known long-rising SNe II. They are most similar in luminosity to PTF09gpn, SN 2005ci, and SN 2009E. At later times ($>$100 days after peak), SN 2018lrq remained brighter than the long-rising SNe II with similarly subluminous peak luminosities.

The remaining five SNe in our sample have peak $r$-band magnitudes around $-16.5$ mag. Specifically, SN 2018lsg peaks at $-16.4$ mag, SN 2020abah peaks at $-16.5$ mag, SNe 2019zfc and 2021wun both peak at $-16.6$ mag, and SN 2019bsw peaks at $-16.7$ mag. SN 1998A is the only long-rising SN II from the literature peaking in this magnitude range, so our CLU sample appears to fill in a (previous) gap in the known long-rising SN II luminosity function. We caution, however, that the ZTF $r$-band is slightly different than the r/R-band photometry used in the older literature sources.

\section{Spectroscopic Properties}
\label{sec:spectroscopy}

\begin{figure*}[!p]
    \centering
    \includegraphics[scale=0.7]{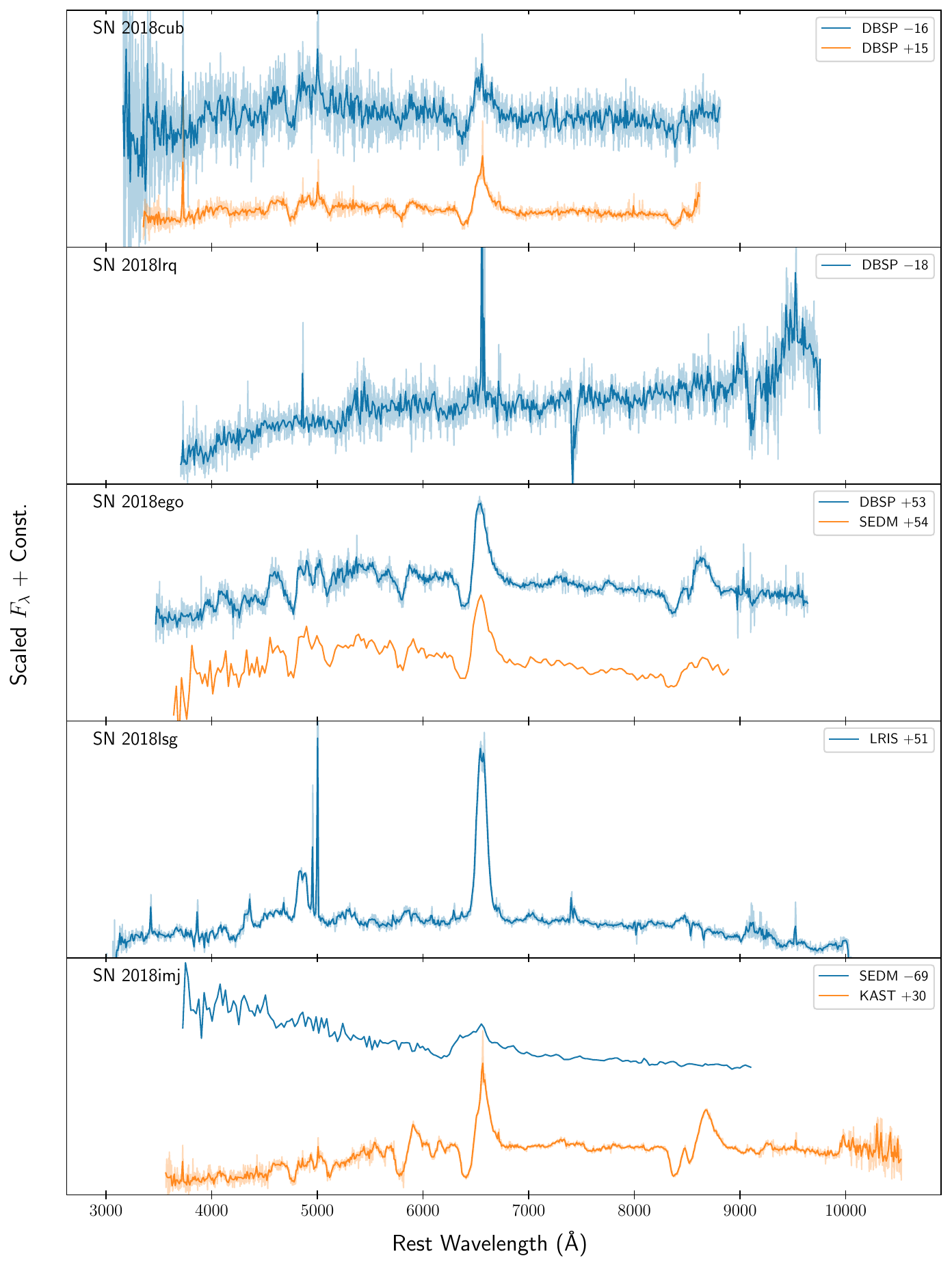}
    \caption{Optical spectra of SNe 2018cub, 2018lrq, 2018ego, 2018lsg, and 2018imj. Spectra were shifted to rest wavelength using the redshifts reported in Table \ref{table: SN info} and scaled by dividing by the median flux. The phase of each spectrum is denoted with respect to the $r$-band peak of the \texttt{sncosmo} fitted light curve. The KAST spectrum of SN 2018imj was obtained from publicly available data on TNS \citep{2018imj_TNS_class}. Higher resolution DBSP and LRIS spectra were binned (10 pixels per bin) to improve signal-to-noise and shown with a darker line; the original spectrum is shown by the lighter colored line of the same color. 
    SEDM, DBSP, and LRIS spectra will be available online at original resolution.}
    \label{fig: spectra 1}
\end{figure*}

\begin{figure*}[!p]
    \centering
    \includegraphics[scale=0.7]{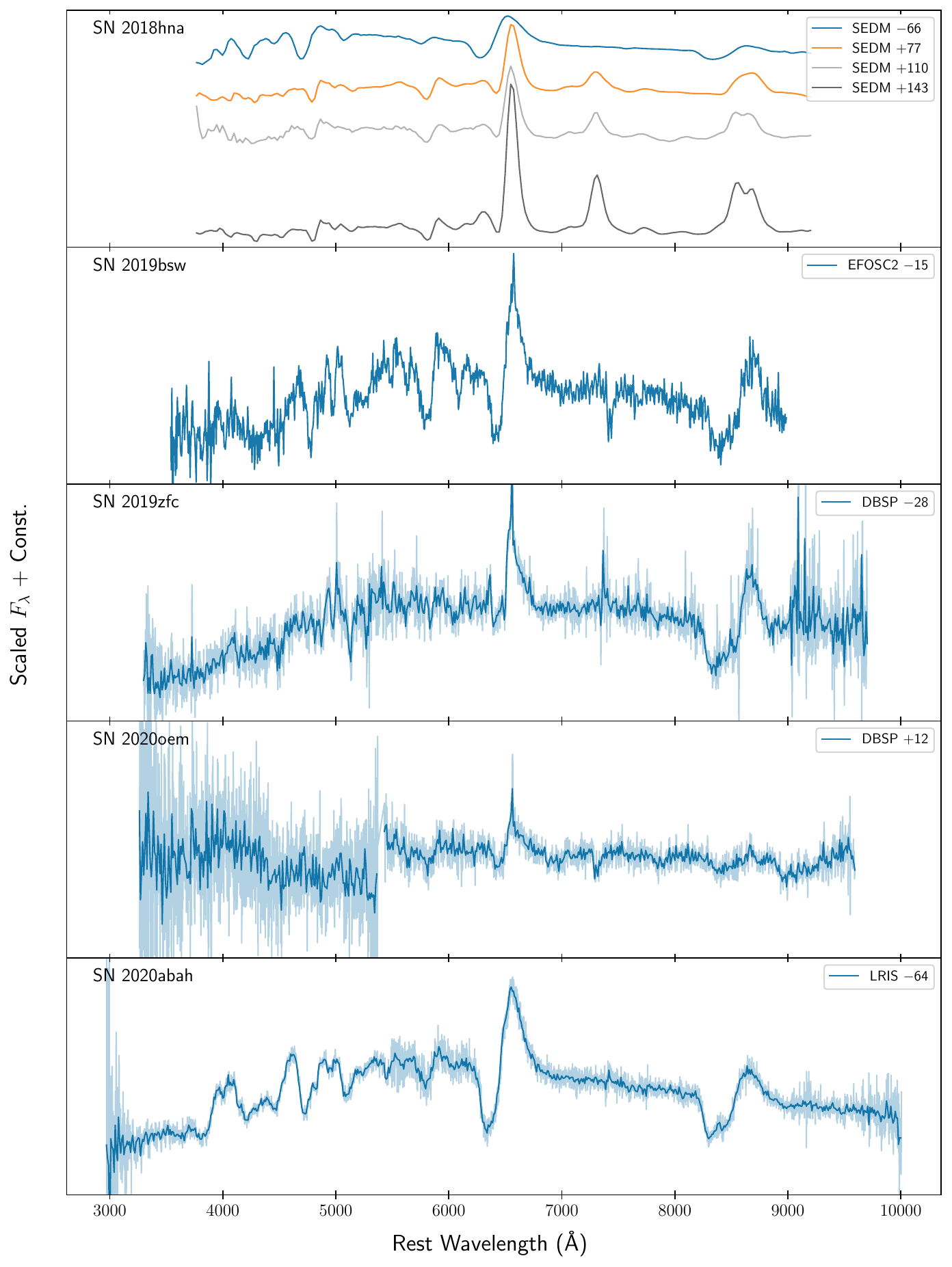}
    \caption{Same as Figures \ref{fig: spectra 1} and \ref{fig: spectra 3} for SNe 2018hna, 2019bsw, 2019zfc, ZTF20abjntg, and 2020abah. For SN 2018hna, only SEDM spectra obtained from regular ZTF classification and followup programs are presented; see \citet{Singh2019} for a more complete optical spectral sequence. The EFOSC2 spectrum of SN 2019bsw was obtained from publicly available data on TNS \citep{2019bsw_TNS_class}. 
    SEDM, DBSP, and LRIS spectra will be available online at original resolution.}
    \label{fig: spectra 2}
\end{figure*}

\begin{figure*}[!t]
    \centering
    \includegraphics[scale=0.7]{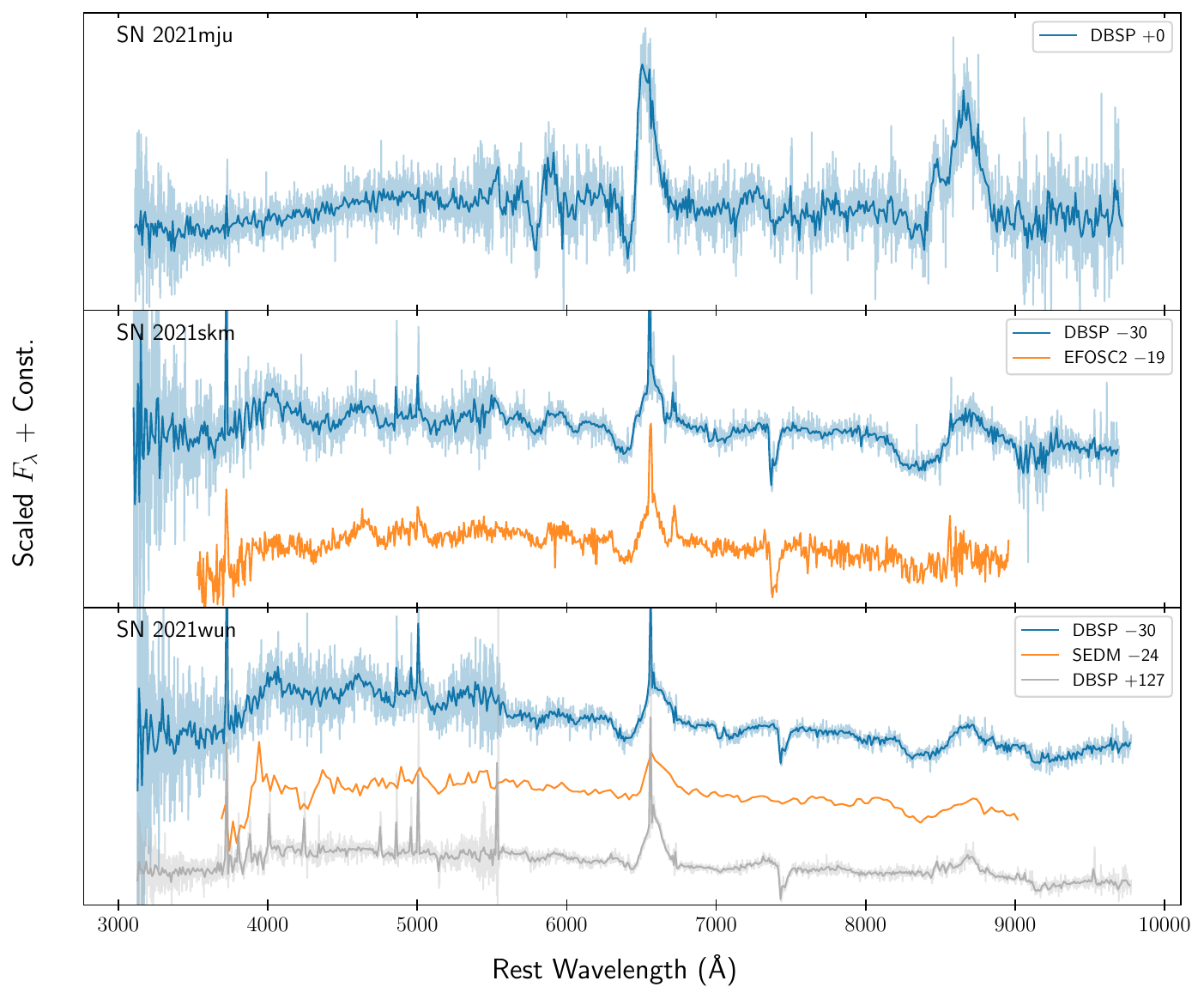}
    \caption{Same as Figures \ref{fig: spectra 1} and \ref{fig: spectra 2} for SNe 2021mju, 2021skm, and 2021wun. The EFOSC2 spectrum of SN 2021skm was obtained from publicly available data on TNS \citep{2021skm_TNS_class}. 
    SEDM, DBSP, and LRIS spectra will be available online at original resolution.}
    \label{fig: spectra 3}
\end{figure*}

\subsection{The CLU Sample of Long-rising SNe II}
In general, spectroscopic followup for transients saved to CLU is intended for classification, and as a result, the majority of our sample only have one to two spectra as no further followup was assigned after classification as a SN II. The instruments regularly used for ZTF CLU classification (see \S\ref{sec:ZTF CLU} for more details) are: the Spectral Energy Distribution Machine (SEDM; \citealt{Blagorodnova2018}) on the Palomar 60-inch telescope, the Double Beam Spectrograph (DBSP; \citealt{DBSP}) on the Palomar 200-inch Hale telescope, and the Low-Resolution Imaging Spectrograph (LRIS; \citealt{LRIS}) on the Keck-I telescope. We supplement our spectra with public data from TNS if available. All spectra were corrected to their rest wavelength using the redshift of their respective host galaxies. The corrected spectra for each SN II in our sample are plotted in Figures \ref{fig: spectra 1}, \ref{fig: spectra 2}, and \ref{fig: spectra 3}. The phase is presented as time from peak, where the peak is estimated from the \texttt{sncosmo} model fit to SN 1987A.

\subsubsection{SN 2018cub / ZTF18aaikcbb}
Two spectra were obtained for SN 2018cub using DBSP, one about 16 days before peak on 2018 June 16 and one about 15 days after peak on 2018 July 17. There does not appear to be significant evolution during the $\approx$30 day period between the spectra; the main difference is the significant improvement in signal-to-noise in the post-peak spectrum. Both spectra show the broad H$\alpha$ P-Cygni feature characteristic of SNe II as well as strong lines at H$\beta$, the \ion{Na}{1} doublet, and the \ion{Ca}{2} triplet. With higher signal-to-noise, \ion{Fe}{2} becomes more obvious, with clear lines at 5169 and 5108 \AA. \ion{Ba}{2} $\lambda$4554,6142 absorption is visible in the post-peak spectrum as well, making SN 2018cub more spectroscopically similar to SN 1987A, which was noted to have unusually strong \ion{Ba}{2} in its spectra \citep{Mazzali1995}.

\subsubsection{SN 2018lrq / ZTF18aaxzlmy}
The single spectrum for SN 2018lrq was taken approximately 18 days prior to peak, on 2018 June 21, with DBSP. The spectrum is highly galaxy-dominated and does not have very high signal to noise. Underlying the narrow host galaxy H$\alpha$ emission is a faint, broad feature that is the SN H$\alpha$. Upon very close inspection, there are possible signs of \ion{Fe}{2} $\lambda$4924, 5018, 5129 and Ca II $\lambda$8498, 8542, 8662. Little further information about the transient can be reliably obtained from this spectrum.

\subsubsection{SN 2018ego / ZTF18ablhrpz}
Two spectra were obtained for SN 2018ego: one using DBSP on 2018 September 12 (53 days after peak) and one using SEDM on 2018 September 13 (54 days after peak). The two spectra are very similar, with the primary difference being the resolution. The P-Cygni shape of H$\alpha$ is very strong in both spectra, and all other Balmer lines (H$\beta$, H$\gamma$, H$\delta$) also show strong absorption. The H$\delta$ absorption is less clear in the lower-resolution SEDM spectrum. \ion{Na}{1} D, the Ca II triplet, and \ion{Fe}{2} multiplet 42 are also strongly seen in both spectra. The separation between the three \ion{Fe}{2} lines (at 4924, 5018, and 5129 \AA) and between Ca II $\lambda$8498, 8542 and Ca II $\lambda$8662 is seen more clearly in the higher-resolution DBSP spectrum. \ion{Ba}{2} is present in the spectra as well, with the 4554\AA\ line much stronger than the 6142\AA\ line. In the SEDM spectrum, the \ion{Ba}{2} $\lambda$6142 line is particularly difficult to see; this is possibly due to the low resolution of the instrument, which can cause the line to blend with H$\alpha$ (e.g., \citealt{Kozyreva2022} Section 3.3.1).

\subsubsection{SN 2018lsg / ZTF18abtjmns}
One spectrum of SN 2018lsg was taken with LRIS on 2018 November 10, about 51 days after peak. The spectrum shows broad H$\alpha$ emission and some weak P-Cygni profiles of H$\beta$, H$\gamma$, and H$\delta$. Unusually, the H$\alpha$ does not exhibit a P-Cygni profile; however, the emission is broad enough to make clear that this was not a missed SN IIn. The Ca II triplet is seen in very narrow absorption lines. This SN occurred in a particularly bright host; the unusually high number of remaining narrow emission lines and the jagged shape of the H$\alpha$ emission peak suggest that there may have been poor background subtraction in the reduction of this spectrum. 

\subsubsection{SN 2018imj / ZTF18abytyif}
There are two spectra available for SN 2018imj. The low-resolution SEDM spectrum was taken on 2018 September 27 (69 days before peak) and shows a very blue continuum with a broad H$\alpha$ and H$\beta$. The second spectrum was taken by the KAST spectrograph on the 3-m Lick telescope about 30 days after peak (on 2019 January 4) and is publicly available on TNS \citep{2018imj_TNS_class}. At a glance, the KAST spectrum strongly resembles that of typical SNe II, with its broad and strong P-Cygni profile for H$\alpha$. The other Balmer lines are not very prominent in the spectrum, partially due to the bluest end having somewhat lower signal-to-noise, although H$\beta$ absorption is still visible. The \ion{Na}{1} D, \ion{Ba}{2} $\lambda$6142, and the Ca II triplet all exhibit strong P-Cygni profiles and there is clear absorption in \ion{Fe}{2} $\lambda$5169,5108.

\begin{figure*}[!p]
    \centering
    \includegraphics[scale=0.7]{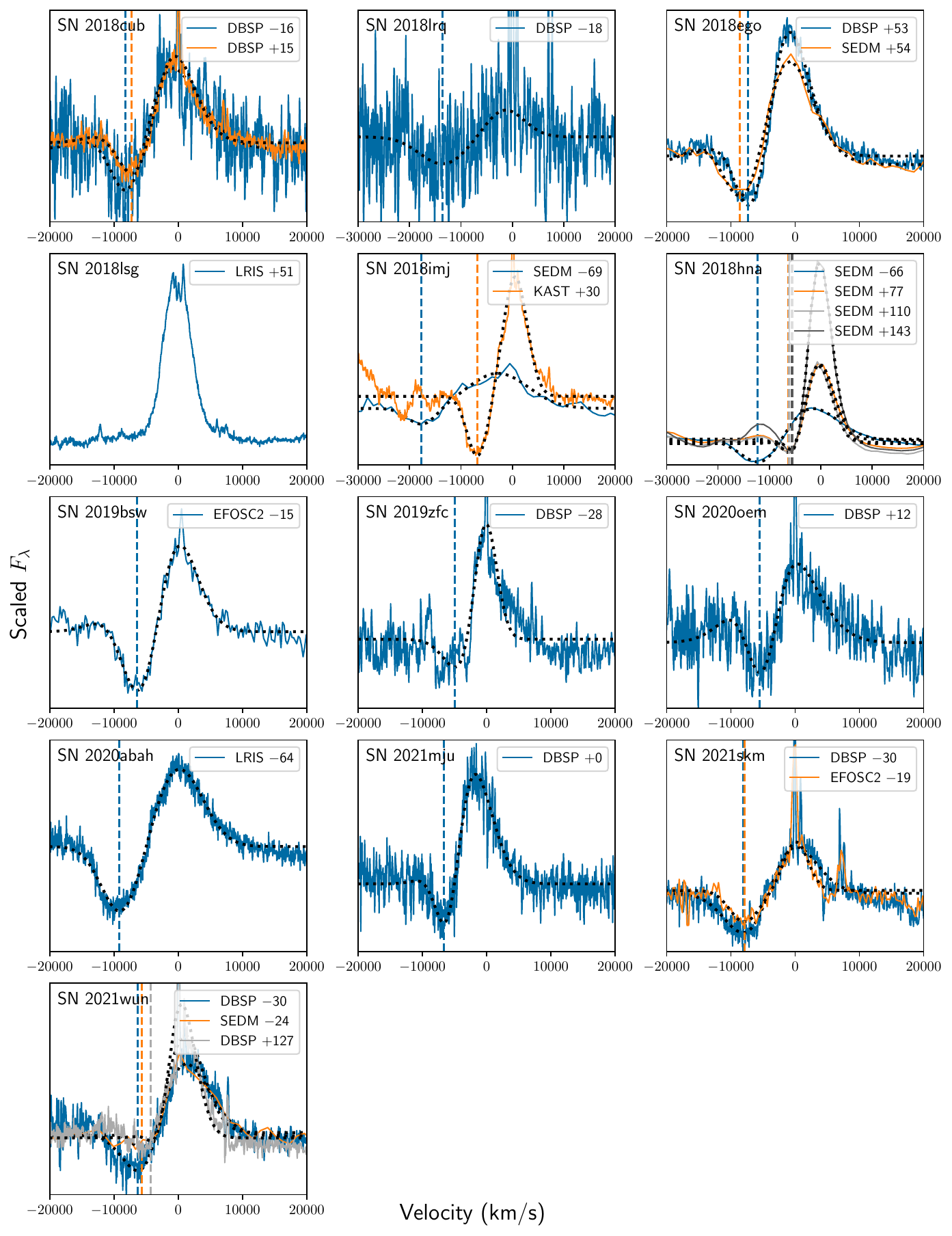}
    \caption{H$\alpha$ line profiles of our CLU sample in velocity space. For each optical spectrum (as presented in Figures \ref{fig: spectra 1}, \ref{fig: spectra 2}, and \ref{fig: spectra 3}), the P-Cygni line profile was fit with a double Gaussian (dotted black lines), and the minimum from the fit is indicated by the dashed line of matching color to the spectrum. The H$\alpha$ profile of SN 2018lsg was not fit due to lack of an absorption feature. The line velocities are plotted as a function of phase in Figure \ref{fig: Ha velocities over time}.}
    \label{fig: Halpha line velocities}
\end{figure*}

\subsubsection{SN 2018hna / ZTF18acbwaxk}
SN 2018hna is the only previously published long-rising SN II in our sample, so we refer to the more complete optical spectral sequences published by \citet{Singh2019} and \citet{Xiang2023}. In this work, we present only the 4 SEDM spectra obtained as part of regular ZTF classification and followup programs. These spectra were obtained on 2018 November 2 (66 days before peak), 2019 March 25 (77 days after peak), 2019 April 27 (110 days after peak), and 2019 May 30 (143 days after peak). In the earliest spectrum, all the optical Balmer lines are visible, with H$\alpha$ and H$\beta$ in particular showing strong P-Cygni profiles. The Ca II triplet, \ion{Na}{1} D, \ion{Fe}{2} $\lambda$5169, and \ion{Ba}{2} $\lambda$4554 also show strong absorption features in this very early spectrum. The next spectrum, taken 77 days after peak, shows the beginning of the transition from the photospheric to the nebular phase. It has significantly lower velocity Balmer lines; only H$\alpha$ appears stronger than in the pre-peak spectrum. \ion{Na}{1} D shows stronger absorption at a lower velocity, and \ion{Ba}{2} $\lambda$6242 as well as additional \ion{Fe}{2} lines have appeared. There is no more absorption at the Ca II triplet; rather, the SN is beginning to show emission features. This spectrum also shows signs of forbidden emission of [O I] and [Ca II]. The next spectrum at 110 days after peak is generally similar to the previous, except for stronger [O I] emission and signs of separation between Ca II $\lambda$8498, 8542 and Ca II $\lambda$8662. The \ion{Ba}{2} $\lambda$6242 absorption and \ion{Fe}{2} $\lambda$5108, 5269 lines are also slightly stronger. The last SEDM spectrum, 143 days after peak, shows significantly stronger H$\alpha$, Ca II, [O I], and [Ca II] emission while its \ion{Na}{1} D absorption remains similar to previous spectra.

\subsubsection{SN 2019bsw / ZTF19aajwkbb}
The only available spectrum of SN 2019bsw is publicly available on TNS. It was taken with the ESO Faint Object Spectrograph and Camera 2 (EFOSC2) on 2019 March 19, 15 days before peak \citep{2019bsw_TNS_class}. Strong P-Cygni profiles are seen for H$\alpha$, \ion{Na}{1} D, and the Ca II triplet, although the lines in the latter are blended. \ion{Fe}{2} $\lambda$4924, 5018, 5169 is also very clearly visible. H$\beta$ and \ion{Ba}{2} $\lambda$6142 have strong absorption lines as well. \ion{Ba}{2} at 4554 \AA\ is not clear compared to 6242 \AA; it is possibly blended with H$\gamma$.

\subsubsection{SN 2019zfc / ZTF20aafezcz}
One DBSP spectrum was taken of SN 2019zfc on 2020 January 27, about 28 days before peak. The spectrum has a prominent H$\alpha$ P-Cygni profile and a strong, broad absorption feature at the Ca II triplet. \ion{Na}{1} D and \ion{Fe}{2} mutiplet 42 also show strong absorption that are narrower compared to hydrogen and Ca II. There is some evidence of \ion{Ba}{2} at 6142 and 4470 \AA\ as well; the shape of their absorption feature is somewhat flat at the bottom compared to the other lines visible in ZTF20afezcz.

\subsubsection{SN 2020oem / ZTF20abjwntg}
One spectrum of SN 2020oem was taken with DBSP on 2020 August 29 (about 12 days after peak). The spectrum shows a P-Cygni H$\alpha$ feature as expected, and absorption at \ion{Na}{1} D is also fairly strong. The Ca II triplet appears to be present but the lines are blended together. The blue side of the spectrum has very low signal-to-noise compared to the red side and it is difficult to identify any spectral features in the noise at wavelengths $\lessapprox5600$\AA.

\subsubsection{SN 2020abah / ZTF20actkutp}
One spectrum of SN 2020abah was obtained with LRIS on 2020 December 12 (64 days before peak). The spectrum shows a strong P-Cygni H$\alpha$ profile, as well as clear absorption in the other optical Balmer lines. \ion{Fe}{2} is also very strong in this spectrum, with clear lines at 4924, 5018, and 5169 \AA. \ion{Na}{1} D and the Ca II triplet both show a strong, broad absorption feature; the individual lines of the Ca II triplet are somewhat blended together. \ion{Ba}{2} is also visible in absorption, with the 4554 \AA\ line being stronger than the 6142 \AA\ line.

\subsubsection{SN 2021mju / ZTF21aasksnl}
One spectrum was obtained of SN 2021mju using DBSP very close to peak on 2021 May 31. The H$\alpha$ P-Cygni profile is clearly visible, as is the \ion{Na}{1} D absorption. The Ca II triplet is also quite strong, and Ca II $\lambda$8662 can be distinguished from the other two lines in the triplet. However, the signal to noise in this spectrum is somewhat low overall, so it is difficult to distinguish other, weaker lines.

\subsubsection{SN 2021skm / ZTF21abjcliz}
We obtained one spectrum of SN 2021skm using DBSP on 2021 August 1, 30 days before peak. A second spectrum taken by EFOSC2 on 2021 August 12, about 19 days before peak, is also publicly available on TNS \citep{2021skm_TNS_class}. Both spectra are fairly similar, with the later spectrum exhibiting a slightly redder continuum and higher SNR at the bluest wavelengths. A strong P-Cygni H$\alpha$ profile, a broad Ca II feature with the triplet lines blended together, and \ion{Na}{1} D absorption are seen at similar velocities in both spectra. \ion{Ba}{2} $\lambda$6142 is also visible in both spectra, and the velocity is lower in the later spectrum. The blue side of both spectra between 4000 and 5500\AA\ show three very broad absorption features; they may correspond to H$\beta$, \ion{Fe}{2} $\lambda$5169, and \ion{Ba}{2} $\lambda$4470 but it is not clear.

\subsubsection{SN 2021wun / ZTF21abtephz}
Two spectra of SN 2021wun were taken using DBSP on 2021 August 31 (30 days before peak) and 2022 February 2 (127 days after peak). One additional spectrum was taken with SEDM on 2021 September 6 (24 days before peak). All spectra show a broad H$\alpha$ feature, with the absorption component of the P-Cygni shape more obvious in the higher resolution DBSP spectra. H$\beta$ absorption is visible in the pre-peak DBSP spectrum, and H$\gamma$ is seen in the SEDM spectrum (H$\gamma$ is not clear in the pre-peak DBSP spectrum due to low SNR). All Balmer lines through H$\delta$ are narrow but visible in the post-peak DBSP spectrum. The Ca II NIR triplet also exhibits a broad P-Cygni-like absorption feature in the pre-peak DBSP and SEDM spectra; the Ca II is beginning to show some emission in the later DBSP spectrum. The pre-peak DBSP spectrum shows \ion{Ba}{2} $\lambda$6142, but \ion{Ba}{2} $\lambda$4470 is more unclear due to the low blue-side SNR. \ion{Ba}{2} is no longer clear in the post-peak spectrum. Both DBSP spectra also have \ion{Na}{1} D absorption, and strong \ion{Fe}{2} $\lambda$5169 absorption is visible in all three spectra.

\begin{figure}[!t]
    \centering
    \includegraphics[scale=0.38]{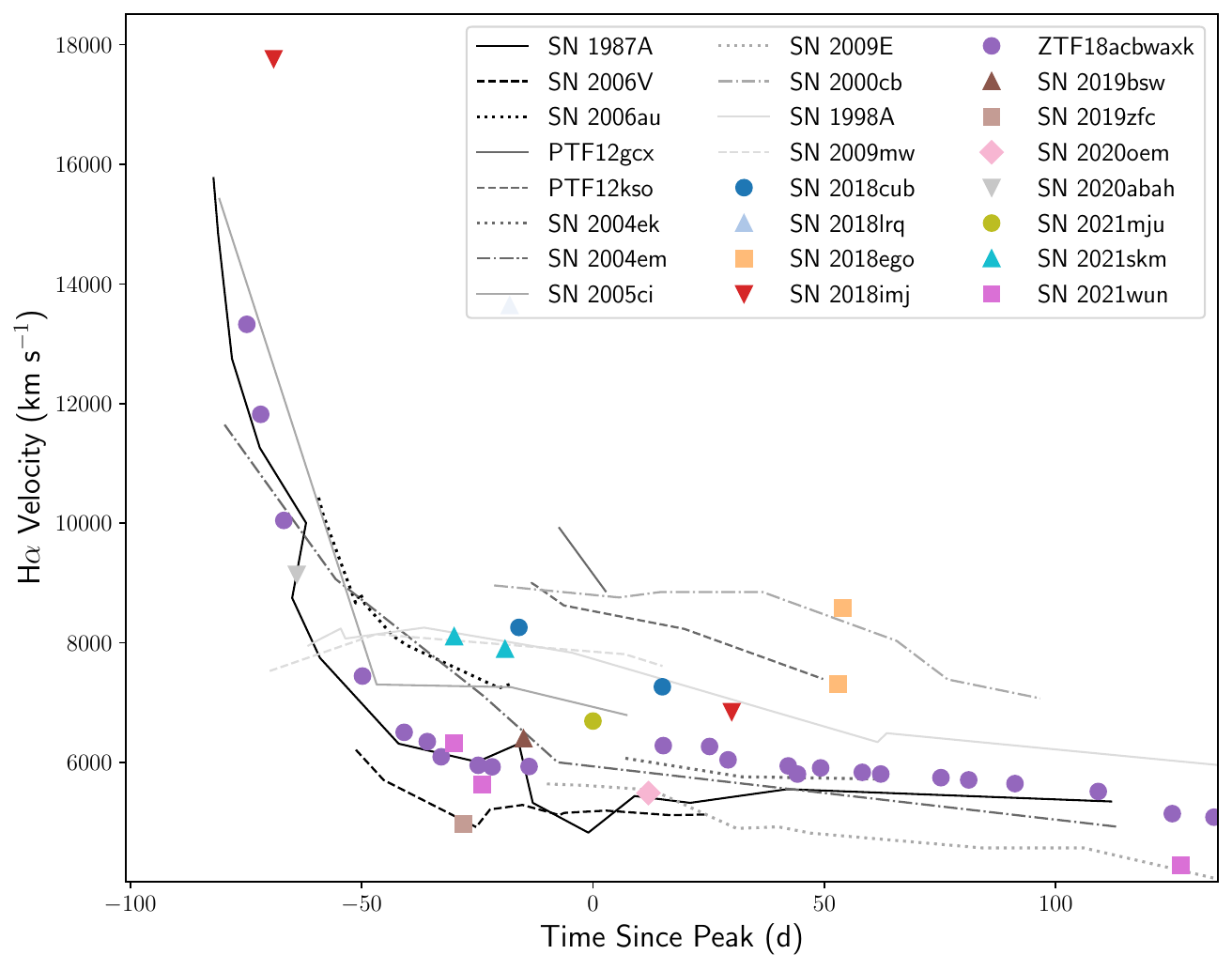}
    \caption{H$\alpha$ velocities as a function of time from peak for our CLU sample (colored points) and other long-rising SNe II from the literature (grayscale lines). Velocity was measured from the P-Cygni line minimum of each spectrum. For SN 2018hna, spectra from the more complete spectral sequence in \citet{Singh2019} were used instead of the SEDM spectra presented in Figure \ref{fig: spectra 2}. Spectra for literature sources obtained via WISeREP \citep{wiserep} from \citealt{1987Aspec} (SN 1987A), \citealt{Taddia2012} (SNe 2006V, 2006au), \citealt{Taddia2016} (PTF sources, SNe 2004ek, 2004em, 2005ci), \citealt{Pastorello2012} (SN 2009E), \citealt{Kleiser2011} (SN 2000cb), \citealt{Pastorello2005} (SN 1998A), and \citealt{Takats2016} (SN 2009mw). Due to the galaxy dominated nature of SN 2018lrq's spectrum, the plotted velocity may not be particularly reliable. SN 2018lsg was excluded from this plot because its H$\alpha$ line did not show any absorption minimum.}
    \label{fig: Ha velocities over time}
\end{figure}

\begin{figure}[!t]
    \centering
    \includegraphics[scale=0.38]{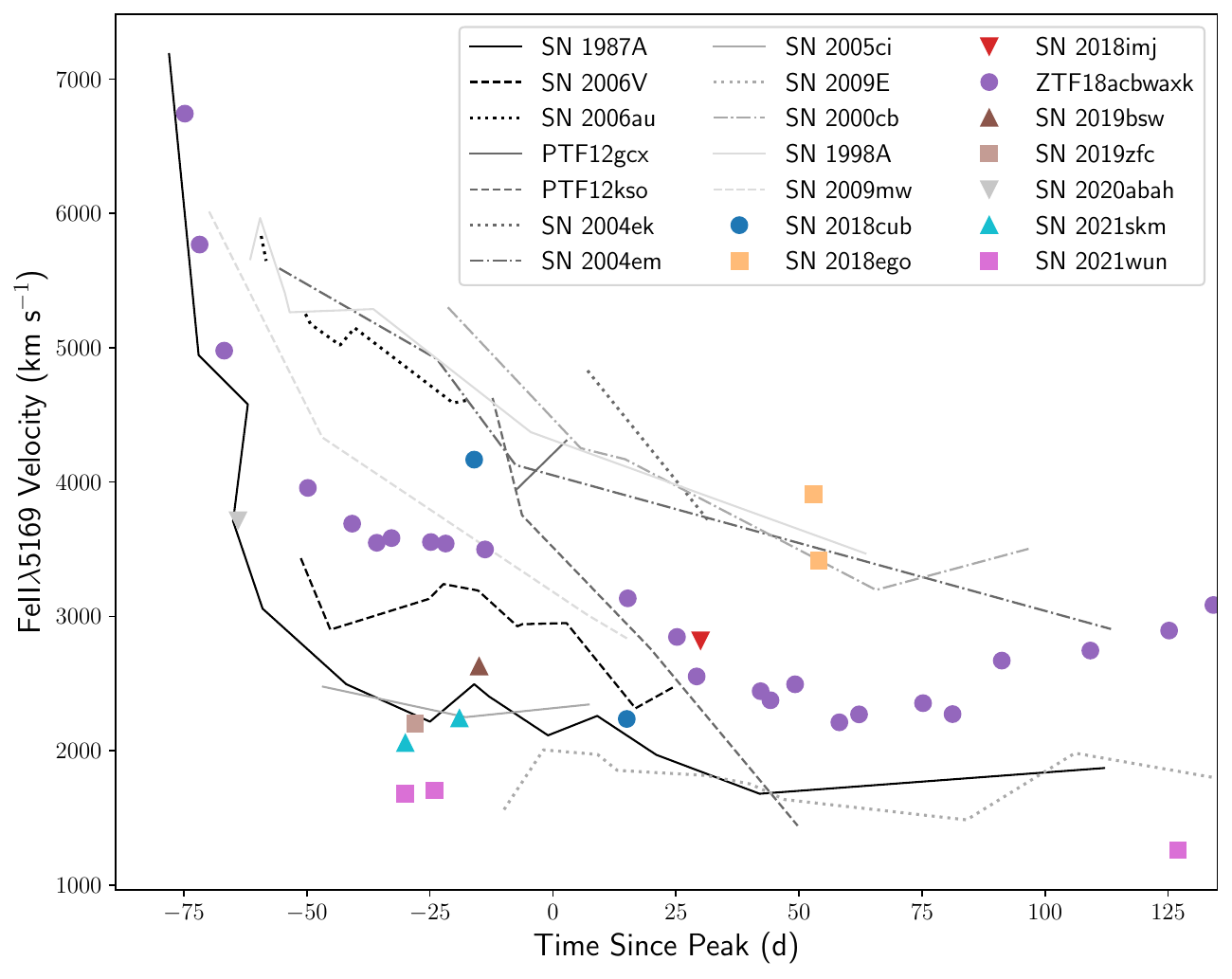}
    \caption{Same as Figure \ref{fig: Ha velocities over time}, but for \ion{Fe}{2}$\lambda$5169 velocities measured from the absorption line minimum. SNe 2018lrq, 2018lsg, 2020oem, and 2021mju were excluded because they did not show a clear absorption feature at \ion{Fe}{2} $\lambda$5169.}
    \label{fig: FeII velocities over time}
\end{figure}

\subsection{Velocities}

All line velocities in this section are calculated from the P-Cygni minimum found via fitting a double Gaussian to the line profile. In cases where the P-Cygni shape is not obvious but there is a clear absorption feature, a single Gaussian was fit to data around the line minimum. In Figure \ref{fig: Halpha line velocities}, we plot the H$\alpha$ line profiles of our CLU sample of long-rising SNe II in velocity space. All events in our sample show broad H$\alpha$ in their spectra, as expected of spectroscopically classified SNe II. All spectra show a P-Cygni profile at H$\alpha$ except SN 2018lsg, which only shows a broad emission feature and no absorption.

The H$\alpha$ velocities of both the literature and our CLU sample are plotted as a function of time in Figure \ref{fig: Ha velocities over time}. In general, the H$\alpha$ velocities of our sample are consistent with the literature. The only outlier is SN 2018lrq, which has a very high velocity of $\approx$13500 km s$^{-1}$ at $\approx$2 weeks prior to peak and is noticeably faster than any other SN in our sample or in the literature at the same phase. However, the spectrum of SN 2018lrq was highly galaxy dominated, so the shape of its H$\alpha$ feature is somewhat unclear. The two spectra of SN 2018ego show fairly different velocities despite being taken a day apart; however, but this is likely explained by \ion{Ba}{2} $\lambda$ blending into the H$\alpha$ at the lower SEDM resolution leading to an apparent higher H$\alpha$ velocity (see \citealt{Kozyreva2022} Section 3.3.1 for more detailed discussion of this effect).

The photospheric velocities of SNe II are better represented by the velocity of \ion{Fe}{2} $\lambda$5169 \citep{Dessart2005}. We measure the \ion{Fe}{2} $\lambda$5169 line velocities in the CLU sample where an absorption feature was visible, and plot the velocities as a function of time in Figure \ref{fig: FeII velocities over time}. The \ion{Fe}{2} $\lambda$5169 velocities of the 9 SNe in our sample we were able to measure are all generally within the expected range of the literature.

\begin{figure*}[!tp]
    \centering
    \includegraphics[scale=0.6]{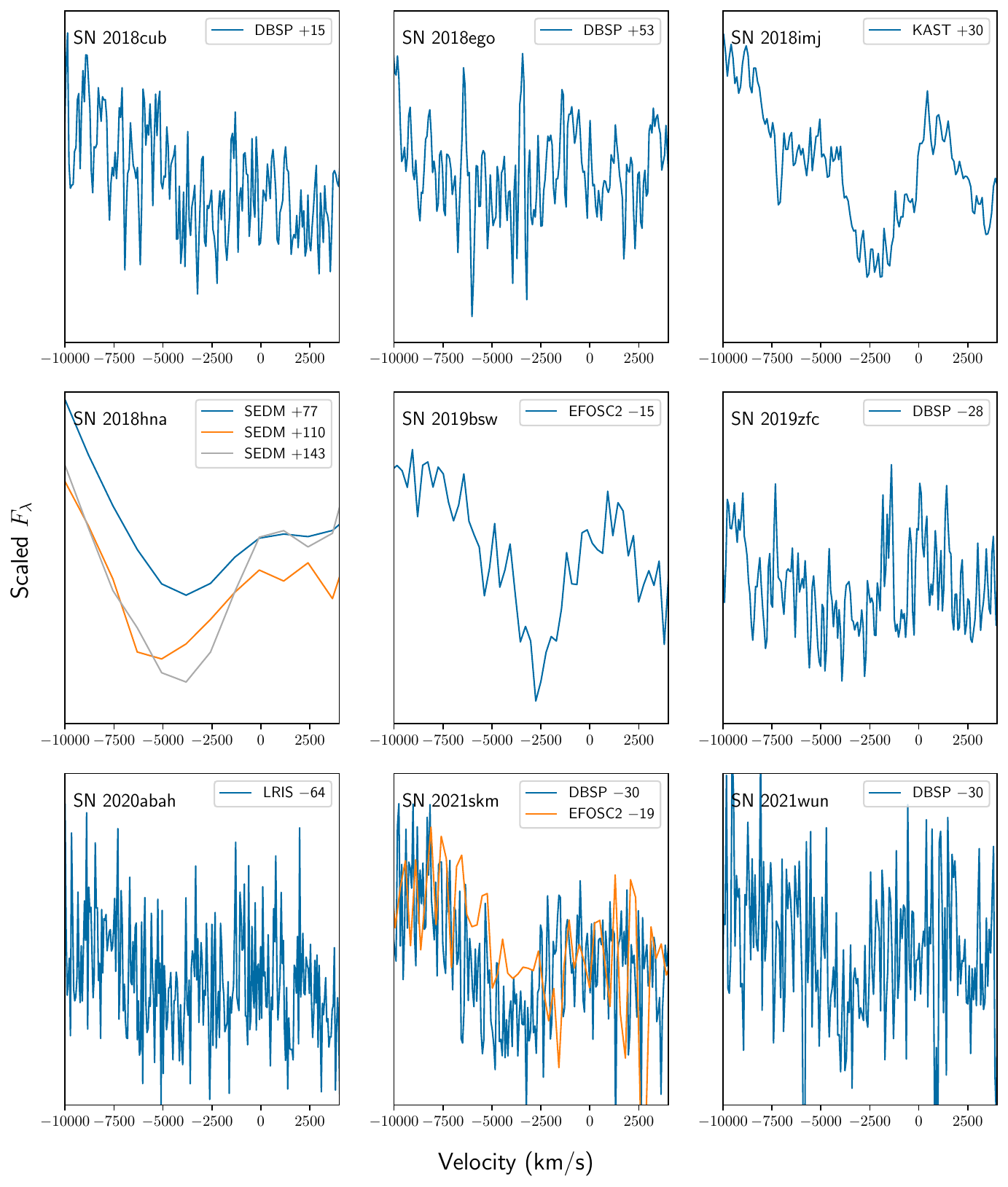}
    \caption{\ion{Ba}{2} $\lambda$6142 line profiles of the 9 SNe with tentative \ion{Ba}{2} detections (absorption visible at either 6242\AA\ or 4554\AA) in velocity space. For events with multiple spectra, spectra with no visible \ion{Ba}{2} absorption were not included.}
    \label{fig: Ba II 6142 line velocities}
\end{figure*}

\begin{figure*}[!tp]
    \centering
    \includegraphics[scale=0.6]{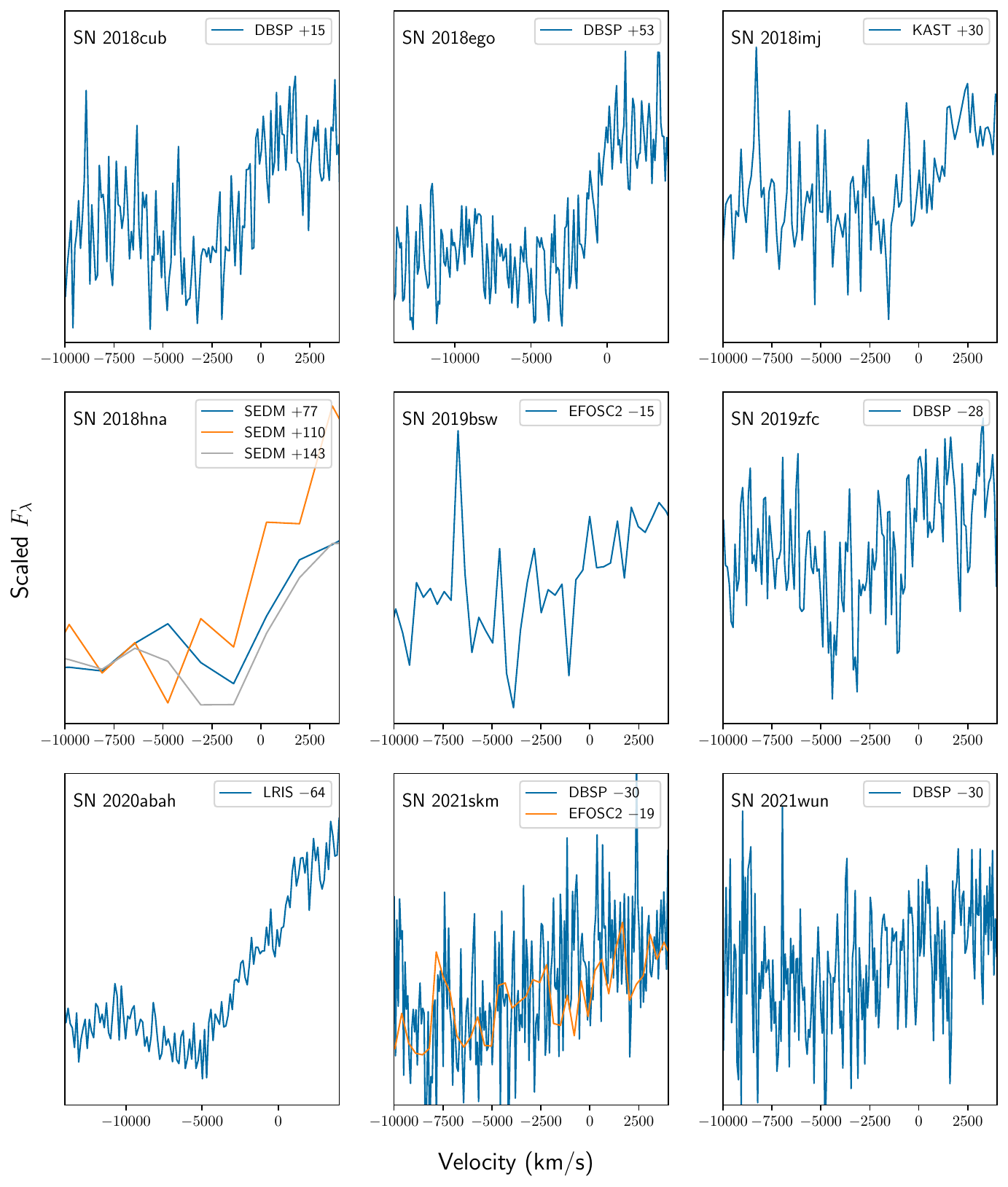}
    \caption{Same as Figure \ref{fig: Ba II 6142 line velocities}, but for \ion{Ba}{2} $\lambda$4454.}
    \label{fig: Ba II 4454 line velocities}
\end{figure*}

The \ion{Ba}{2} $\lambda$6142 line is also a good indicator of photospheric velocity, and the presence of \ion{Ba}{2} is considered characteristic of 1987A-like events \citep{Mazzali1995}. We examine the \ion{Ba}{2} $\lambda$4454, 6242 lines in our sample of CLU events and tentatively identify \ion{Ba}{2} absorption in 9 out of 13 sources. The 6142\AA\ and 4454\AA\ line profiles are plotted for these 9 SNe in Figures \ref{fig: Ba II 6142 line velocities} and \ref{fig: Ba II 4454 line velocities} respectively. SNe 2018imj, 2018hna, 2019bsw, and 2021skm have the strongest \ion{Ba}{2} $\lambda$6142 lines; absorption is also somewhat visible in the other five sources but are more obscured by noise. SNe 2018cub, 2018ego, 2018hna, and 2020abah have the strongest \ion{Ba}{2} $\lambda$4454 lines. While SNe 2018imj, 2019bsw, and 2021skm both showed very strong \ion{Ba}{2} $\lambda$6142, their 4454\AA\ line is notably weaker or nonexistent/obscured by low signal-to-noise in that region of the spectrum. SN 2021wun is similar, except the BaII $\lambda$6142 line is noisier than the other three. The opposite is true of SN 2018ego and (to a somewhat lesser degree) SNe 2018cub and 2020abah. SN 2019zfc has noisy but visible absorption features at both lines.

\section{Host Galaxy Properties}
\label{sec:hosts}

\begin{figure*}[!t]
    \centering
    \includegraphics[width=\textwidth]{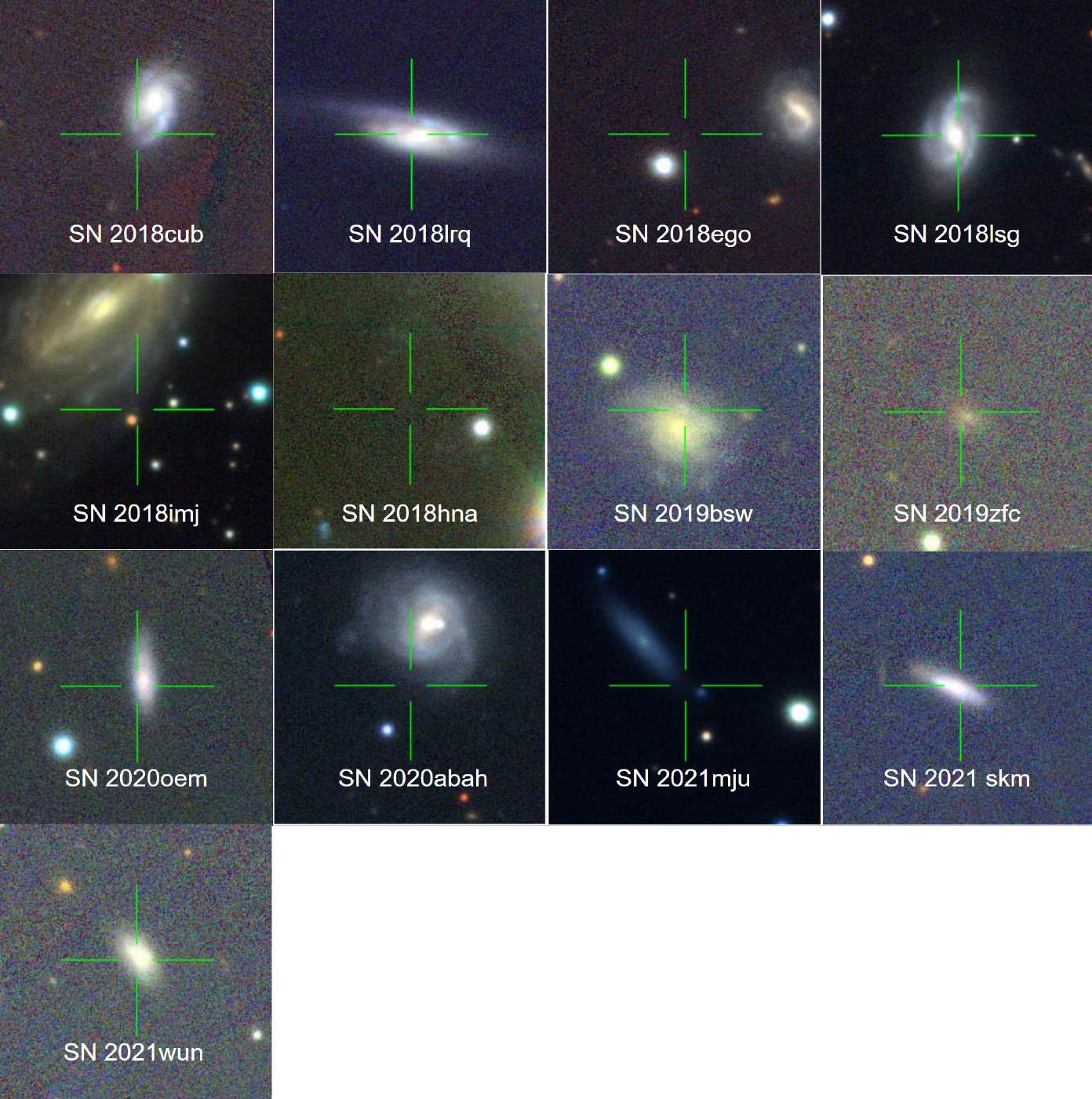}
    \caption{Archival images of the locations of the long-rising SNe II in our ZTF CLU sample. Images are taken from the PS1 survey \citep{Chambers2016} and the cutouts are 60$\arcsec\times$60$\arcsec$ in size. The green crosshairs indicate the location of the SN.}
    \label{fig: host galaxy images}
\end{figure*}

In Figure \ref{fig: host galaxy images}, we plot the location of all 13 long-rising SNe in our ZTF CLU sample relative to their host galaxies.

SN 1987A-like SNe have been observed to occur primarily in low metallicity environments on the outskirts of bright late-type galaxies or dwarf galaxies \citep{Pastorello2012,Taddia2013}. We investigate this trend for the CLU sample using both an indirect estimation method (\S\ref{subsec: indirect metallicity est}) and direct measurement from galaxy spectra using the N2 emission line diagnostic (\S\ref{subsec: direct metallicity measurement}). We compare our results for both methods to those presented in \citet{Taddia2013}'s detailed study of 1987A-like SNe host galaxies in Figure \ref{fig: host metallicity comparison}.

\begin{deluxetable*}{cccccccc}[!tb]
\tablewidth{0pt}
\tablecaption{Host Galaxy Properties of CLU Long-rising SNe II \label{table: host galaxies}}
\tablehead{
\colhead{} & \colhead{} & \colhead{$\alpha$ (J2000)}\vspace{-0.4cm} & \colhead{$\delta$ (J2000)} & \colhead{Major Axis ($2R_{25}$)} & \colhead{Minor Axis ($2b$)} & \colhead{PA} & \colhead{} \\
\colhead{SN Name} & \colhead{Host Galaxy} & \colhead{}\vspace{-0.4cm} & \colhead{} & \colhead{} & \colhead{} & \colhead{} & \colhead{$r_{SN}/R_{25}$}\\
\colhead{} & \colhead{} & \colhead{(hh:mm:ss)} & \colhead{(dd:mm:ss)} & ($\arcsec$) & ($\arcsec$) & \colhead{($^\circ$)} & \colhead{}}
\startdata
SN 2018cub & WISEA J150541.98+604751.4 & 15:05:41.983 & +60:47:51.57 & 29.06 & 19.18 & 160 & 0.57 \\
SN 2018lrq & MCG+06-30-045 & 13:34:51.236 & +34:03:19.86 & 58.37 & 16.34 & 77 & 0.067 \\
SN 2018ego & 2MASX J15525218+1958107 & 15:52:52.203 & +19:58:10.92 & 29.02 & 24.38 & 19 & 2.16 \\
SN 2018lsg & CGCG 374-012 & 20:46:44.606 & -01:22:07.95 & 29.80\tablenotemark{a} & 15.50 & 0 & 0.0084\\
SN 2018imj & IC 0454 & 06:51:06.30 & +12:55:19.3 & 104.27\tablenotemark{b} & 54.72 & 140 & 0.82 \\
SN 2018hna & UGC 07534 & 12:26:08.10 & +58:19:21.0 & 157.80\tablenotemark{b} & 125.29 & ... & ... \\
SN 2019bsw & WISEA J100506.20-162425.1 & 10:05:06.200 & -16:24:25.80 & ... & ... & ... & ... \\
SN 2019zfc & SDSS J034655.55+000225.9 & 03:46:55.559 & +00:02:26.00 & 8.86 & 7.89 & 63 & 0.37 \\
SN 2020oem & WISEA J152729.72+034646.8 & 15:27:29.725 & +03:46:46.37 & 26.24 & 9.45 & 1 & 0.30 \\
SN 2020abah & CGCG 127-002 & 11:36:53.166 & +21:00:15.14 & 43.24 & 32.86 & 41 & 0.84 \\
SN 2021mju & WISEA J164148.29+192203.6 & 16:41:48.278 & +19:22:03.44 & 43.89 & 11.41 & 42 & 0.64 \\
SN 2021skm & 2MASX J16165615+2148359 & 16:16:56.146 & +21:48:35.56 & 26.64 & 9.86 & 63 & 0.18 \\
SN 2021wun & WISEA J154631.96+252545.7 & 15:46:31.945 & +25:25:45.58 & 17.82 & 10.34 & 34 & 0.0081
\enddata
\tablecomments{Right ascension $\alpha$ and declination $\delta$ are for the host galaxy centers. SN coordinates are reported in Table \ref{table: SN info}. $r_{SN}/R_{25}$ is the deprojected, normalized SN distance from the host galaxy center. Galaxy coordinates, major and minor axes, and position angles taken from NED. The 25 mag SDSS $r$-band isophotal radius is used as the major axis $2R_{25}$ unless otherwise noted. 
\tablenotetext{a}{20 mag 2MASS $K$-band isophotal radius.}
\tablenotetext{b}{25 mag $B$-band isophotal radius}}
\end{deluxetable*}

\subsection{Indirect Metallicity Measurements}
\label{subsec: indirect metallicity est}
We estimate the central metallicity of the host via the $r$-band absolute magnitude following the equation presented in \citet[][Section 2.5]{Arcavi2010}. The absolute magnitude is calculated with the same method as in \S\ref{sec:photometry}, using the known galaxy redshift. The literature \citep[e.g., ][]{Pilyugin2004,Pilyugin2007} has shown that \ion{H}{2} regions have lower metallicities farther away from the center of their galaxy. Following \citet{Taddia2013}, we correct the central metallicity estimate to the SN location. First, we calculate the deprojected, normalized distance of the supernova to the galaxy center $r_{SN}/R_{25}$ following \citet{Hakobyan2009}; the data for these calculations are presented in Table \ref{table: host galaxies}. The central metallicity is then extrapolated to the SN location assuming an average metallicity gradient of $-0.47$ dex $R_{25}^{-1}$ \citep{Pilyugin2004}. We report the estimated metallicities at the host center and the SN location in Table \ref{table: indirect metallicity estimates}.

The directly measured mean metallicity of 1987A-like environments from \citet{Taddia2013} is $12+\log{\textrm{(O/H)}} = 8.36\pm0.05$ dex. Their mean estimated metallicity is slightly higher, at $8.51\pm0.04$ dex. Within this $\approx$0.15 dex range, we find 5 SNe in  our CLU sample (SNe 2018cub, 2018lrq, 2020oem, 2020abah, and 2021skm) generally consistent with these mean estimates. SNe 2018imj, 2021mju, and 2021wun have lower estimated metallicites compared to the mean and the lowest estimate from \citet{Taddia2013} (8.21 dex). However, they are not as metal-poor as their lowest directly measured metallicity (7.96 dex) so these 3 SNe can still be considered consistent, if on the metal-poor side, with the known long-rising SN II host environments. 

The estimated metallicities of SNe 2018ego and 2019zfc are $\gtrapprox$0.2 dex lower than any previously known 1987A-like SN. The low metallicity estimate of SN 2018ego can be explained by its location in the far outskirts of its host galaxy (which is bright and not particularly metal-poor), while SN 2019zfc is explained by its very faint dwarf host. On the other hand, SN 2018lsg has the highest estimated metallicity in our CLU sample although it is still consistent with the most metal-rich 1987A-like SNe in \citet{Taddia2013}; it is most comparable to SN 1998A, which had a directly measured metallicity of $12+\log{\textrm{(O/H)}} = 8.68\pm0.06$ dex. However, SN 1998A was located far from the center of its host, unlike SN 2018lsg; the high metallicity of SN 1998A was explained by its host having an unusually shallow metallicity gradient.

Due to a lack of diameter and/or position angle data on NED, we were unable to calculate $r_{SN}/R_{25}$ and extrapolate the central metallicity estimate to the SN location for SNe 2018hna and 2019bsw. SN 2018hna has a very faint dwarf galaxy as a host, so it is likely to also be one of the more metal-poor 1987A-like SNe. The estimated central metallicity of SN 2018bsw is fairly typical of the bright long-rising SN II hosts, but the SN is located somewhat close to its center on the sky (4\farcs66) and so would likely have a higher than average metallicity for this subclass.

\begin{deluxetable*}{cccccccc}[!tb]
\tablewidth{0pt}
\tablecaption{Estimated Host Environment Metallicity for our CLU Long-rising SNe II \label{table: indirect metallicity estimates}}
\tablehead{
\colhead{} & \colhead{Host $m_r$}\vspace{-0.4cm} & \colhead{Galactic $A_r$} & \colhead{Host $M_r$} & \colhead{$Z/Z_\odot$} & \colhead{$12+\log{\textrm{(O/H)}}$} & \colhead{$Z/Z_\odot$} & \colhead{$12+\log{\textrm{(O/H)}}$}\\
\colhead{SN Name} & \colhead{}\vspace{-0.4cm} & \colhead{} & \colhead{} & \colhead{} & \colhead{} & \colhead{} \\
\colhead{} & \colhead{(mag)} & \colhead{(mag)} & \colhead{(mag)} & (Center) & (Center) & \colhead{(SN)} & \colhead{(SN)}}
\startdata
SN 2018cub & 15.57 & 0.040 & $-20.9$ & 1.01 & 8.69 & 0.55 & 8.43 \\
SN 2018lrq & 15.02 & 0.026 & $-20.2$ & 0.78 & 8.58 & 0.73 & 8.55 \\
SN 2018ego & 15.64 & 0.120 & $-20.4$ & 0.83 & 8.61 & 0.08 & 7.60 \\
SN 2018lsg & 14.44$^\dag$ & 0.156 & $-20.7$ & 0.93 & 8.66 & 0.92 & 8.65 \\ 
SN 2018imj & 13.12$^\dag$ & 0.523 & $-20.2$ & 0.77 & 8.57 & 0.32 & 8.19 \\ 
SN 2018hna & 14.42 & 0.026 & $-15.6$ & 0.15 & 7.86 & -- & -- \\ 
SN 2019bsw & 15.69$^\dag$ & 0.117 & $-19.8$ & 0.77 & 8.57 & -- & -- \\ 
SN 2019zfc & 19.13 & 0.356 & $-16.3$ & 0.18 & 7.95 & 0.12 & 7.77 \\ 
SN 2020oem & 16.72 & 0.116 & $-19.6$ & 0.62 & 8.48 & 0.45 & 8.34 \\ 
SN 2020abah & 14.33 & 0.054 & $-21.6$ & 1.33 & 8.81 & 0.54 & 8.42\\ 
SN 2021mju & 15.99 & 0.175 & $-19.3$ & 0.57 & 8.45 & 0.28 & 8.14 \\ 
SN 2021skm & 16.60 & 0.185 & $-18.9$ & 0.49 & 8.38 & 0.41 & 8.30 \\
SN 2021wun & 17.28 & 0.107 & $-17.7$ & 0.31 & 8.18 & 0.31 & 8.17
\enddata
\tablecomments{Magnitudes are in the SDSS $r$-band except those marked with a dagger ($^\dag$), whose magnitudes are in the PS1 $r$-band. The Galactic extinction $A_r$ was obtained using the NED extinction calculator, which uses the dust maps in \citet{Schlafly2011} and the extinction law of \citet{Fitzpatrick1999} assuming $R_v=3.1$. We assume a solar metallicity of $12+\log{\textrm{(O/H)}}_\odot=8.69$ dex \citep{Asplund2009}. Metallicities near the SN location are extrapolated from the deprojected, normalized distances presented in Table \ref{table: host galaxies}. See \S\ref{subsec: indirect metallicity est} for more details on the metallicity estimation procedures.
}
\end{deluxetable*}

\subsection{Direct Metallicity Measurements} \label{subsec: direct metallicity measurement}
For five SNe in our sample (SNe 2018lrq, 2019zfc, 2020oem, 2021skm, and 2021wun), the available SDSS spectrum of their host galaxy are sufficient to directly measure the metallicity at the SN location. These 5 SNe have $<$3$\arcsec$ (the aperture size of a single fiber on the SDSS spectrograph) of separation from the center of their respective host galaxy. Using the reported SDSS emission line fits, we calculate the N2 emission line diagnostic N$2 = \log{\left([\textrm{\ion{N}{2}}]\lambda 6584 / \textrm{H}\alpha\right)}$. From N2, we obtain the oxygen abundance using the polynomial relation presented in \citet{Pettini2004}: $$12+\log{\textrm{(O/H)}} = 9.37 + 2.03(\textrm{N}2) + 1.2(\textrm{N}2^2) + 0.32(\textrm{N}2^3)$$

Assuming an uncertainty of 0.2 dex in the direct N2 measurement \citep{Taddia2013}, we find that most of these SNe direct measurements broadly consistent with the earlier indirect estimate. SN 2018lrq has a directly measured metallicity of  $12+\log{\textrm{(O/H)}} = 8.63$, consistent with solar metallicity and confirming that SN 2018lrq is a rare example of a high metallicity 1987A-like supernova. SN 2019zfc has a directly measured metallicity of $12+\log{\textrm{(O/H)}} = 8.15$, higher than the indirect metallicity estimate considering the uncertainty. Nonetheless, the host environment of SN 2019zfc remains noticeably metal-poor compared to the rest of our sample and 1987A-like SNe as a whole; in fact, only SN 1909A ($12+\log{\textrm{(O/H)}} = 7.96$ dex, \citealt{Taddia2013}) has a more metal-poor host environment. SNe 2020oem, 2021skm, and 2021wun are all consistent with the other low-metallicity 1987A-like events, with directly measured $12+\log{\textrm{(O/H)}}$ values 8.49, 8.48, and 8.24 respectively.

\begin{figure*}[!t]
    \centering
    \includegraphics[scale=0.6]{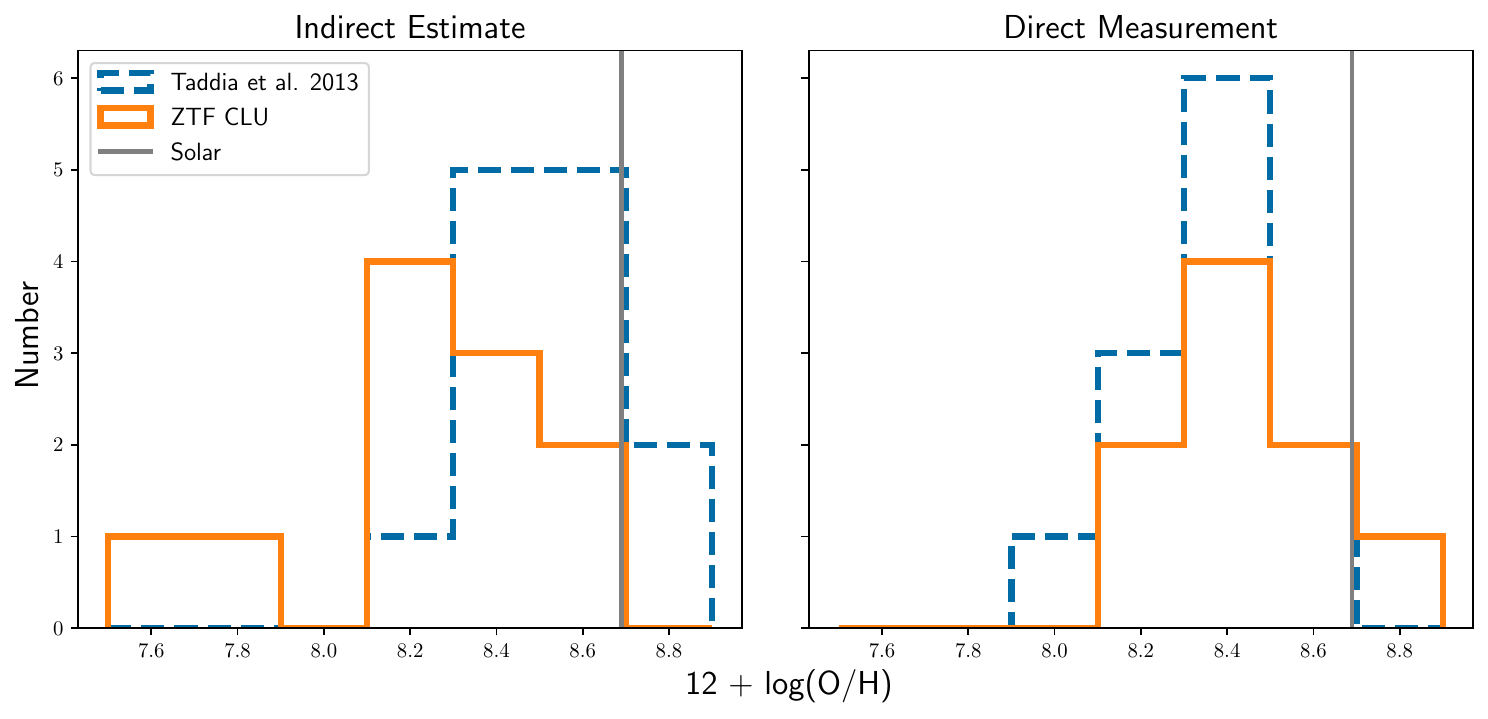}
    \caption{Distribution of the $12+\log{\textrm{(O/H)}}$ at the SN location of our ZTF CLU sample (solid orange lines) and the literature (dashed blue lines). In the left figure, the metallicities are indirectly estimated from host brightness and an assumed metallicity gradient (\S\ref{subsec: indirect metallicity est}; \citealt{Taddia2013} Table 3, Column 3). In the right figure, metallicities are directly measured from galaxy spectra (\S\ref{subsec: direct metallicity measurement}; \citealt{Taddia2013} Table 3, Column 5). The solid gray line indicates solar metallicity, $12+\log{\textrm{(O/H)}}_\odot=8.69$ dex \citep{Asplund2009}.}
    \label{fig: host metallicity comparison}
\end{figure*}

SN 2018ego is too offset from its host galaxy to obtain a meaningful spectrum at its location after the transient faded in order to obtain the host environment metallicity. Notably, this is the only SN in our sample with $r_{SN}/R_{25}>1$. A SDSS spectrum of the center of the host galaxy is available, however, so following the procedure described previously to measure the N2 diagnostic and oxygen abundance, we find a central host metallicity of $12+\log{\textrm{(O/H)}} = 8.64$. Extrapolating this measured central metallicity to the SN location assuming the same metallicity gradient as before, we slightly revise our previous estimate to $12+\log{\textrm{(O/H)}} = 7.79$ dex. However, we do not consider this revised number a true direct measurement and exclude it from Figure \ref{fig: host metallicity comparison}.

SN 2019bsw is not in the SDSS fields, but its host galaxy has an optical spectrum available on NED from the 6dF Galaxy Survey. The 6dF fiber aperture is 6$\arcsec$, which is greater than the 4\farcs66 separation between the SN location and host center, so we can use this spectrum to directly measure the metallicity. We fit the continuum around the H$\alpha$ and [\ion{N}{2}] lines and subtract this continuum from the spectrum. Line fluxes for the H$\alpha$ and [\ion{N}{2}]$\lambda$6584 emission lines were measured using the \texttt{specutils} package \citep{specutils} using careful selection of the relevant line regions. The N2 diagnostic was then calculated from the line fluxes. We find a metallicity of $12+\log{\textrm{(O/H)}} = 8.72$, which is super-solar, and makes SN 2019bsw the highest-metallicity 1987A-like SN known to date.

The host of SN 2018lsg, similarly, has an available 6dF spectrum on NED. The SN is very nuclear so we would be able to use this spectrum to directly measure the host metallicity. However, the 6dF spectrum shows an unexpected broad emission feature around H$\alpha$ which may be indicative of an AGN flare at the time of observation. As a result, we were unable to adequately measure the H$\alpha$ and [\ion{N}{2}] lines from this spectrum.

For 3 of the remaining SNe in our CLU sample, we observed their host galaxies with Keck I/LRIS. For each source, the long slit mask was angled to include both the galaxy nucleus and the location of the SN. Using \texttt{lpipe} \citep{Perley2019}, a spectrum of the galaxy center and \ion{H}{2} regions falling into the slit near the SN location were obtained. For each spectrum, N2 was measured from the H$\alpha$ and [\ion{N}{2}]$\lambda$6584 lines using the same method described above for SN 2019bsw.

IC 0454, the host of SN 2018imj, was observed on the night of 2022 February 2. From the spectrum extracted from the galaxy center, we measure $12+\log{\textrm{(O/H)}} = 8.69$, $\approx$0.1 dex higher than the previous estimate. We were able to extract a second spectrum from the closest \ion{H}{2} region to the SN location, $\approx$25$\arcsec$ and measured a metallicity of $12+\log{\textrm{(O/H)}} = 8.47$ (0.60$Z_\odot$). We take this number as the directly measured estimate of the host metallicity of SN 2018imj.

Host spectra of SNe 2018cub and 2021mju were obtained on the night of 2022 July 27. We measure $12+\log{\textrm{(O/H)}} = 8.74$ at the center of SN 2018cub's host galaxy using the LRIS spectrum. The closest \ion{H}{2} region to the SN location where we could extract a spectrum is $\approx$7$\arcsec$ away from the center, and we measure a metallicity of $12+\log{\textrm{(O/H)}} = 8.47$. Our central metallicity measurement is 0.06 dex higher than the central metallicity as measured using SDSS reported line fluxes of the same galaxy, so we correct the \ion{H}{2} region metallicity to $12+\log{\textrm{(O/H)}} = 8.41$ (0.52$Z_\odot$) and take this number as the directly measured host metallicity of SN 2018cub. We measure a metallicity of $12+\log{\textrm{(O/H)}} = 9.02$ at the center of the host of SN 2021mju using LRIS. The metal-rich nature of this host is supported by the central metallicity of the galaxy as measured by SDSS line fluxes ($12+\log{\textrm{(O/H)}} = 8.91$) being 0.11 dex lower than our LRIS measurement. The closest \ion{H}{2} region to the SN location where we could extract a spectrum with no contribution from a nearby background galaxy was $\approx$11$\arcsec$ from the center. We measured $12+\log{\textrm{(O/H)}} = 8.64$ at this location. After correcting for the difference between the central LRIS and SDSS measurements, we take $12+\log{\textrm{(O/H)}} = 8.53$ as the directly measured host metallicity of SN 2021mju. 

We were unable to obtain LRIS host spectra for SNe 2018hna and 2020abah.

In Figure \ref{fig: host metallicity comparison}, we compare metallicity values between the \citet{Taddia2013} sample and our CLU sample of both the indirect estimation and direct measurement methods described in \S\ref{table: indirect metallicity estimates} and \S\ref{subsec: direct metallicity measurement}. In general, our sample is consistent with previously described long-rising SN II environments with subsolar metallicities. However, we do have one super-solar source, SN 2019bsw. We also find that unlike \citet{Taddia2013}, whose indirect estimates were generally overestimates of their directly measured metallicities, our indirect estimation values were underestimates of the direct measurements.

\section{Rate of Long-rising SN II Events}
\label{sec:rates}

We estimate the volumetric rates of long-rising type II/1987A-like supernovae using the \texttt{simsurvey} code \citep{simsurvey}. \texttt{simsurvey} simulates light curves as they would be observed given a light curve template (using \texttt{sncosmo}) and survey plan (including observation times, filters, and sky brightness).  As presented in Figures \ref{figures: gold sample full LCs 1} and \ref{figures: gold sample full LCs 2}, the built-in SN 1987A model in \texttt{sncosmo} is generally a good fit to the shape of all light curves in our sample, and it was used as a qualitative check of the photometry-based selection criteria. Therefore, the built-in SN 1987A \texttt{sncosmo} model was used as the light curve template. The actual ZTF observing history between 2018 June 1 and 2021 December 31 was used as the survey plan.

Using our controlled and systematic sample of ZTF CLU events, we can derive a luminosity function of long-rising SNe II. Due to the small overall sample size ($N=13$), we do not distinguish between the different CLU parameters of ZTF-I and ZTF-II. The ZTF-II CLU experiment set a luminosity limit ($M\geq-17$) that did not exist in ZTF-I, and thus there may be a selection bias against any luminous long-rising SNe II occurring after 2021 January 1. However, all ZTF-I SNe in our sample with $M<-17$ occurred outside of the smaller ZTF-II volume, so this effect should be negligible. Our sample size of CLU events is as large as the existing number of known long-rising SNe II in the literature, so to avoid introducing any additional selection biases to the luminosity function we do not include any additional sources from the literature sample. 

\begin{figure}[!t]
    \centering
    \includegraphics[scale=0.47]{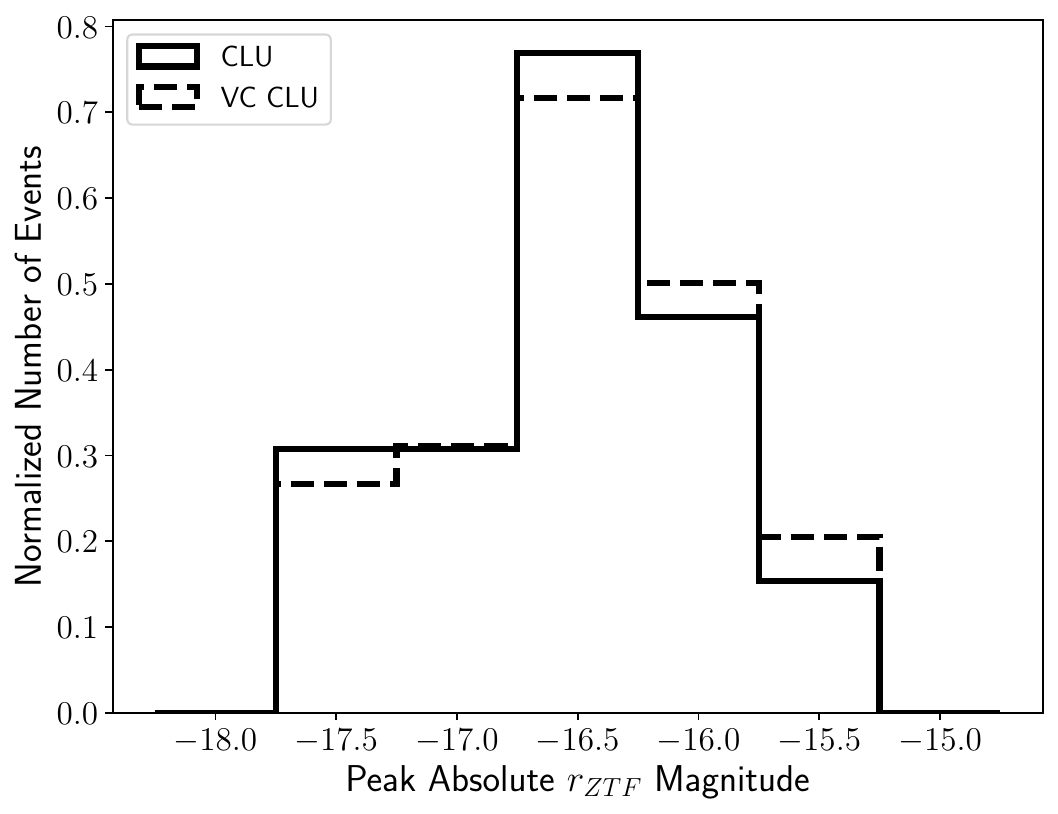}
    \caption{Peak $r$-band absolute magnitude distribution of our ZTF-CLU sample of long-rising SNe II. The luminosity function was volume corrected by applying a $1/V_{max}$ weighting to each object (see text for details); this volume-corrected function is shown by the dashed line.}
    \label{fig: lumfunc}
\end{figure}

A volume correction was applied to the luminosity function by weighing each object by $1/V_{max}$, where $V_{max}$ is the maximum distance out to which a transient could be detectable out to a limiting magnitude of 20. For sources in ZTF-I detectable past 200 Mpc and sources in ZTF-II detectable past 150 Mpc, $V_{max}$ was set to the maximum CLU experiment volume. The luminosity function, both before and after volume correction, is presented in Figure \ref{fig: lumfunc}.

\begin{figure*}[!t]
    \centering
    \includegraphics[width=0.49\textwidth]{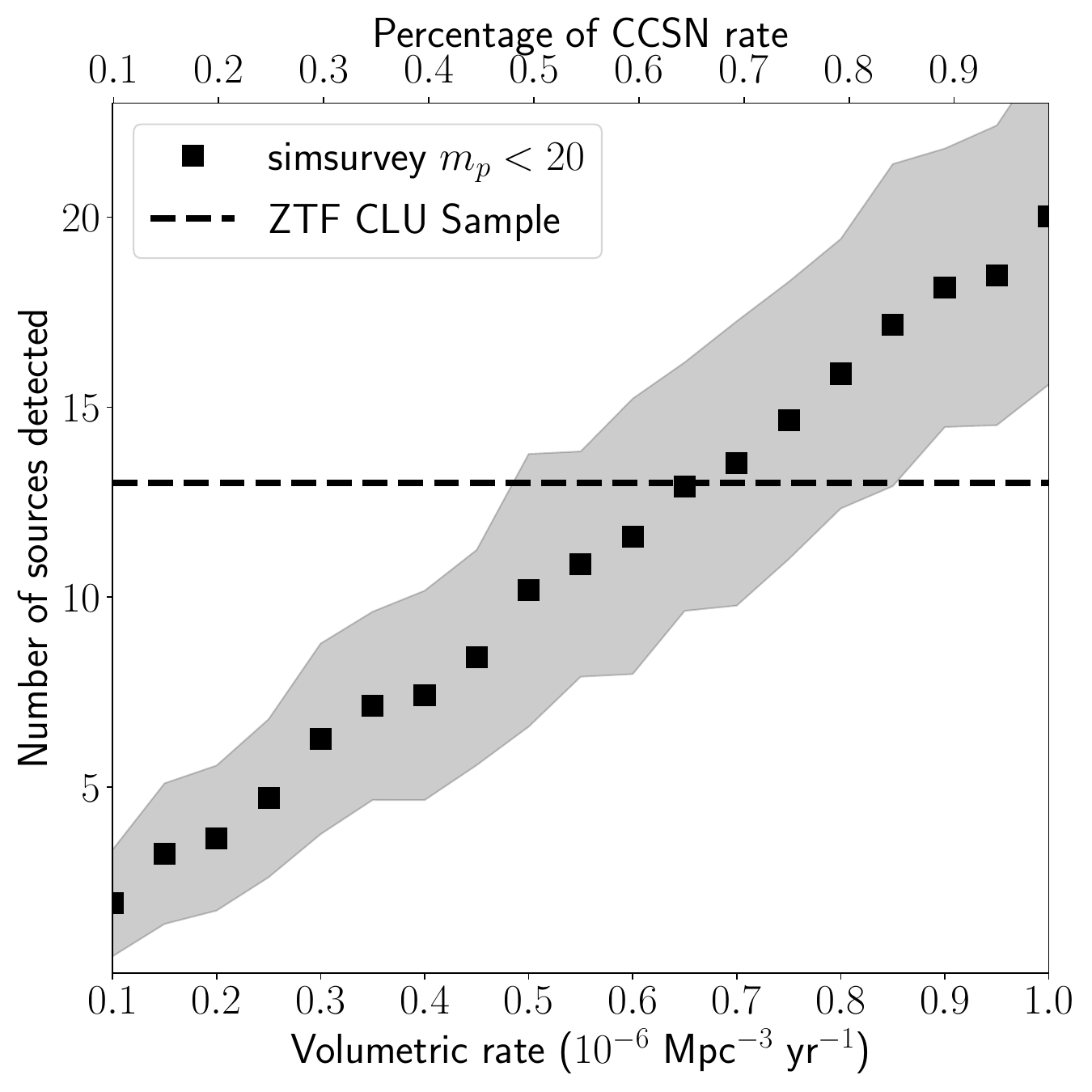}
    \includegraphics[width=0.49\textwidth]{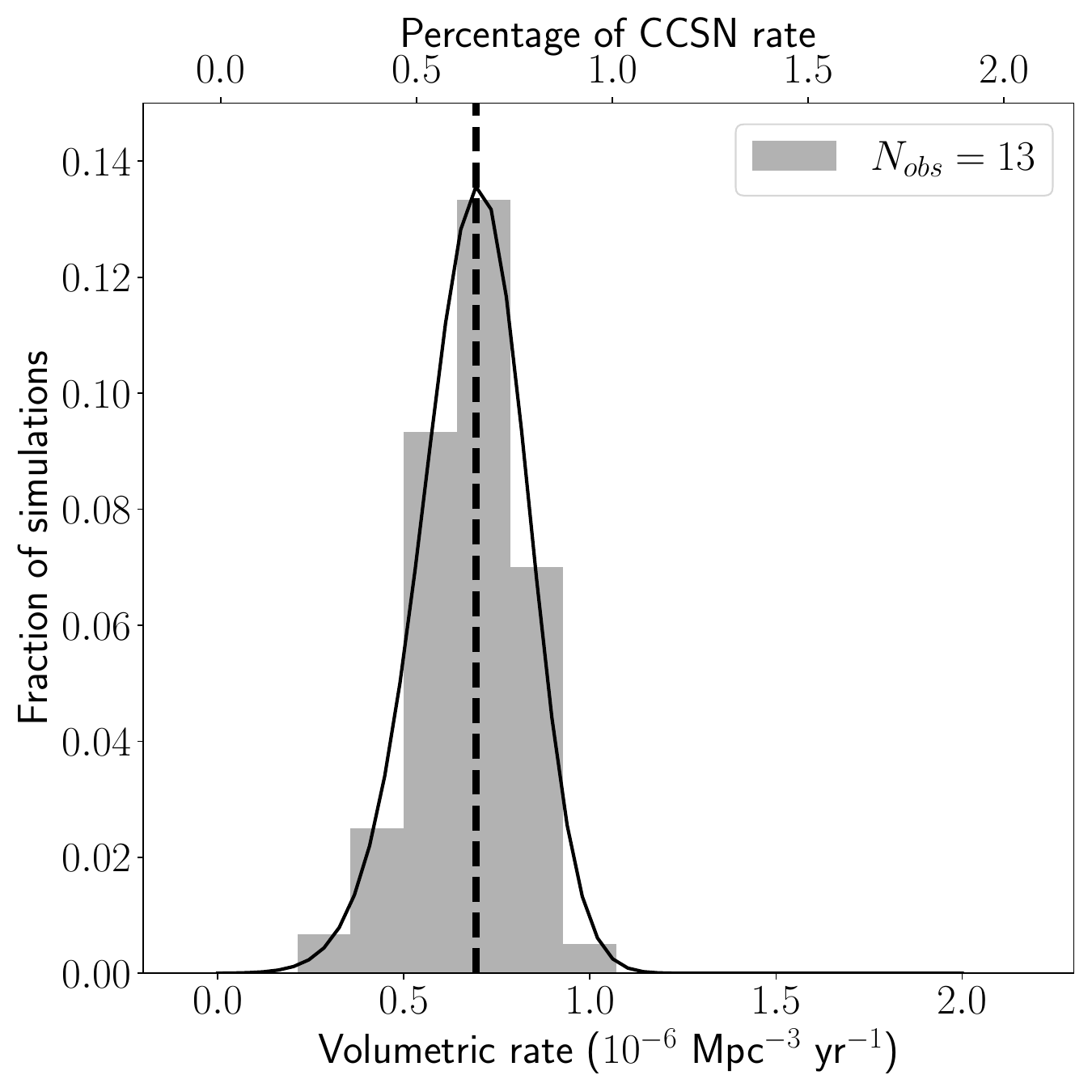}
    \caption{Estimations of the long-rising SNe II rate using \texttt{simsurvey}. On left, we present the number of 1987A-like sources ``detected", ie. passing the filtering function, as a function of the volumetric rate. The points and error bars show the mean and standard deviation of the number of detected transients, calculated from 100 iterations of the simulated survey at each input volumetric rate. The dashed black line shows the number of confirmed long-rising SNe II in our ZTF CLU sample. On right, we present the fraction of simulations with the observed number of transients as a function of the volumetric rate. In both figures the assumed CCSN rate is $10.1\times\ 10^4\ \textrm{Gpc}^{-3}\ \textrm{yr}^{-1}$ \citep{Perley2020}.}
    \label{fig: 87A rates}
\end{figure*}

To select simulated long-rising SNe II from \texttt{simsurvey}, we must define a filtering function that the simulated light curves will be passed through. We reuse the forced photometry-based selection criteria described in \S\ref{subsec:SN II sample}) as the basis of this function. Then, we add the following criteria to define the final \texttt{simsurvey} filtering function:

\begin{enumerate}
    \item There must be at least one 5$\sigma$ detection prior to peak and at least two 5$\sigma$ detections brighter than the median survey limiting magnitude of 20.5 mag. The peak apparent magnitude (in the $r$- or $g$-bands) must also be brighter than 20 mag. These requirements ensure that the transient is detected by the ZTF alert system to be passed to the CLU filter.
    \item The transient must have $z<0.05$ if the first detection falls in the ZTF-I time period (prior to 2021 January 1) or $z<0.0331$ if the first detection is during the ZTF-II time period (2021 January 1 and after). If the transient is in ZTF-II, it must also have an absolute magnitude $M\geq-17$, inclusive of error bars (e.g., a peak magnitude of $17.1\pm0.1$ would be included in the sample). This requirement ensures that the transient would have been saved to the CLU filter and assigned appropriate spectroscopic follow-up.
\end{enumerate}

Applying this final filter function to the forced photometry light curves (as presented in \S\ref{sec:photometry}) returns all 13 sources in our sample.

We simulate ZTF light curves of long-rising SNe II over a range of input volumetric rates. The simulated transients were injected by \texttt{simsurvey} out to a redshift of $z=0.05$ following the luminosity function in Figure \ref{fig: lumfunc}. At each rate, 100 simulations of the survey plan were executed. For each simulation, the number of transients that passed the filtering function was recorded. We estimate the number of detected transients at each rate by taking the average number of detected transients over the 100 simulations, and use the standard deviation as the uncertainty. Figure \ref{fig: 87A rates} shows the expected number of detected transients and the fraction of simulations producing our observed sample size as a function of the input volumetric rate.

The distribution of the fraction of simulations matching our sample size is well fitted with a skewed normal distribution. This functional form was used to derive the rate and uncertainty using a 68\% confidence interval, and we find a volumetric rate of
$$ \left(6.94_{-1.53}^{+1.30}\right) \times 10^{-7}\ \textrm{Mpc}^{-3}\ \textrm{yr}^{-1}$$

This raw \texttt{simsurvey} rate already accounts for the luminosity cut implemented in the ZTF-II CLU experiment because the cut was included in the filtering function used to select the simulated light curves. As we used the real observing history of ZTF to simulate the light curves as they would have been observed, \texttt{simsurvey} also accounts for survey conditions such as weather and cadence.

\begin{figure}[!t]
    \centering
    \includegraphics[scale=0.55]{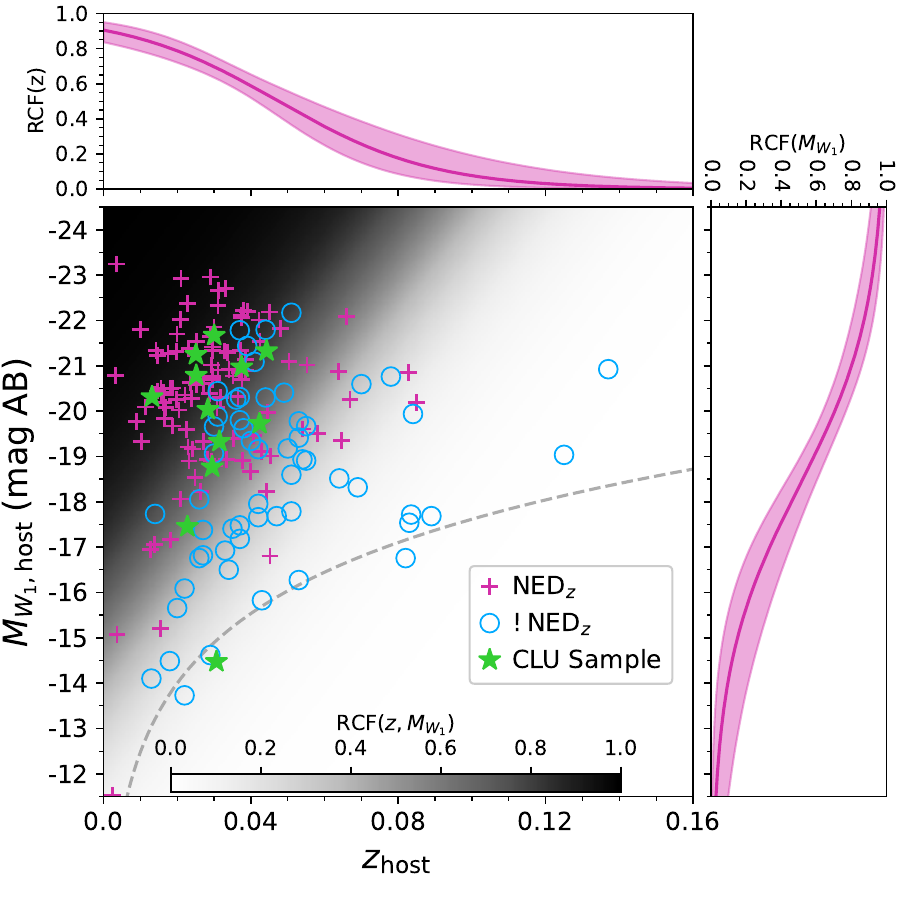}
    \caption{Absolute WISE $W_1$ band magnitude vs redshift for host galaxies of the 161 CCSNe in the ZTF BTS sample classified between 2018 April 1 and 2018 December 31 \citep{Fremling2020}. The magenta pluses represent galaxies with known redshift via NED or similar databases, and the blue circles represent galaxies without previously known redshift. The background shading in the central plot represents $RCF(z, M_{W_1})$, the probability of a host galaxy having a known redshift based on its redshift $z$ and absolute WISE $W_1$ magnitude $M_{W_1}$. The dashed line shows the WISE $W_1$ detection limit $M_{W_1, limit}\approx20.699$ mag \citep{Schlafly2019}. The probability of the host galaxy having a known redshift based only on its redshift ($RCF(z)$, top plot) or $W_1$ absolute magnitude ($RCF(M_{W_1})$, right plot) are shown by the magenta curves, where the solid line represents the median and the shaded region represents the 90\% credible interval. Green stars indicate where the host galaxies of our ZTF CLU sample of long-rising SNe II fall in the RCF.}
    \label{fig: CCSNe RCF}
\end{figure}

By design, we can only identify supernovae in the CLU experiment that occur in galaxies with a known spectroscopic redshift. Therefore, we must correct the \texttt{simsurvey} rate of 87A-events for galaxy catalog incompleteness. We use the redshift completeness factor (RCF), or the probability that a random galaxy will have a known spectroscopic redshift \citep{Kulkarni2018}, from the ZTF Bright Transient Survey (BTS) \citep{Fremling2020}. The RCF is a function of host galaxy redshift and the host galaxy mass, for which the WISE $W_1$ (3.36 $\mu$m) absolute magnitude is used as a proxy. The BTS sample of Ia SNe used to derive the RCF in \citet{Fremling2020} because these SNe occur in both star-forming and passive galaxies; in contrast, CC SNe trace only star-forming galaxies. Since long-rising type II SNe discussed in this work are a subset of CC SNe, however, we instead follow the same procedure to re-derive the RCF using only CC SNe in the BTS sample. This result is presented in Figure \ref{fig: CCSNe RCF}.

For the long-rising SNe II in the CLU sample, we obtain the redshift and WISE $W_1$ magnitude of each host galaxy. The $W_1$ magnitudes were obtained from \texttt{Tractor} SDSS/WISE forced photometry \citep{Lang2016}. There were three sources, SNe 2018lsg, 2018imj, and 2019bsw, that were not located in the SDSS fields and therefore have no \texttt{Tractor} forced photometry available for their host galaxies. In this case, the profile-fit $W_1$ magnitude reported on NED was used instead. For each host galaxy, we calculate $RCF(M_{W_1}, z)$. Then, we compute the harmonic mean of the RCF values and find that the incompleteness of galaxy catalogs leads to an underestimate of the 1987A-like rate of $\approx$49\%. SN 2018hna's host galaxy UGC 07534 has absolute $W_1$ magnitude $M_{W_1}=-10.8$, fainter than the faintest BTS CC SNe host galaxy used to derive the RCF, so it was excluded from the calculation. The tendency for many long-rising SNe II to occur in the low-metallicity environments of very faint dwarf galaxies likely contributes to the fairly high RCF correction.

Correcting for the RCF, we find a volumetric rate of long-rising SNe II of
$$ \left(1.37_{-0.30}^{+0.26}\right) \times 10^{-6}\ \textrm{Mpc}^{-3}\ \textrm{yr}^{-1}$$

This rate is $1.4\pm0.3$\% of the total core-collapse supernova rate in the local universe ($1.01\times\ 10^{-4}\ \textrm{Mpc}^{-3}\ \textrm{yr}^{-1}$, \citealt{Perley2020}), consistent with previous estimates of the rate of long-rising SNe II in the literature. \citet{Taddia2016} estimates that long-rising SNe II comprise $\sim$1\% of all CCSNe within a distance modulus of $\mu\approx35$ by counting CCSNe observed by the Palomar Transient Factory, and \citet{Pastorello2012} estimates $<$1.5\% of SNe II within the same volume using the Asiago Supernova Catalogue. Older studies estimated slightly higher rates for such events: $\sim$2\% \citep{Kleiser2011} or $<$3\% \citep{Smartt2009} of all CCSNe. Notably, all of these previous estimates of the rate of BSG explosions assume a smaller volume than our ZTF CLU sample, which includes transients detected out to a volume of $z=0.05$ ($\mu\approx36.8$) in ZTF-I and $z=0.0331$ ($\mu\approx35.9$) in ZTF-II. Therefore, we have expanded the considered volume up to $\sim$100 Mpc farther than previous rate estimates for similar events.

From June 2018 to December 2020, the ZTF-I CLU experiment achieved an overall completeness rate of 76.8\%, with 1711 out of 2229 transients having spectroscopic classifications. Of these classified transients, 773 (45.2\%) were SN II and 10 were long-rising. In 2021, the ZTF-II CLU experiment achieved an overall completeness rate of 76.4\%, spectroscopically classifying 191 out of 250 extragalactic transients. Of the classified transients, 125 (65.4\%) were SN II; of those SNe II, 3 were long-rising. Overall, we expect to have missed no more than 3 long-rising SN II in 2.5 years of the ZTF-I CLU experiment and $<$1 in the first year of ZTF-II. In general, the long-lived nature and relative ease of identifying a H$\alpha$ P-Cygni feature in spectra should make 1987A-like events difficult to miss due to spectroscopic incompleteness.

The efficiency of the ZTF image subtraction pipeline is affected by transient magnitude and underlying surface brightness. Our sample of long-rising SNe II exhibit a range of apparent magnitudes (see \S\ref{sec:photometry}) and occur in a variety of host environments (see \S\ref{sec:hosts}); for example, SN 2018lsg is located close to the bright nucleus of its host galaxy while SN 2018ego is extremely far from its host nucleus. The subtraction efficiency directly affects the likelihood of each transient detection. We have not corrected for the ZTF detection efficiency, so our rate is conservatively a lower limit. Thus, we present the first lower limit of long-rising SNe II derived from a large, systematic, volume-limited experiment. 

\section{Summary and Discussion}
\label{sec:discussion}
In this work, we presented a sample of 13 long-rising SNe II systematically selected from the Census of the Local Universe experiment of the Zwicky Transient Facility. With 15 previously known long-rising SNe II other than SN 1987A in the local universe ($z\lessapprox0.05$) \citep{Pastorello2005,Kleiser2011,Pastorello2012,Taddia2012,Taddia2016,Takats2016,Singh2019}, this new CLU sample nearly doubles the number of known SNe in this subclass. We find that our sample is generally consistent with previously known SN 1987A-like SNe from the literature with respect to photometric, spectroscopic, and host galaxy properties. Using our sample, we derived the first volumetric rate of long-rising SN II events from a large systematic experiment. We briefly summarize our methods and findings below:

\begin{itemize}
    \item The ZTF Census of the Local Universe is a volume-limited supernova experiment aiming to spectroscopically classify all SNe occurring in galaxies known to be within 200 (before January 2021) or 150 (after January 2021) Mpc. ZTF CLU is well suited for systematic searches of subluminous transient events such as SNe with blue supergiant progenitors similar to SN 1987A ($M_r\approx-16.3$).
    \item Using systematic photometry-based criteria, we identified 13 long-rising SNe II in our sample from a total of 3444 transients saved to the ZTF CLU experiment between June 2018 and December 2021. All SNe in our sample are spectroscopically classified SNe II with $>$40 days of rise and dome-shaped light curves in $r$-band photometry.
    \item The long-rising SNe in our sample exhibit a range of peak absolute magnitudes from $-15.6$ to $-17.5$ in the $r$ band, which is consistent with the range of peak absolute magnitudes observed in the literature. In addition, we ``fill in" a small gap in the previously known long-rising SN II luminosity function at $M_r\approx-16.5\pm0.2$ mag.
    \item As expected of SNe II, all our spectra show broad H$\alpha$ at velocities mostly consistent with those observed in the literature. We also detect common lines such as \ion{Fe}{2} $\lambda$5169, \ion{Na}{1} D, and the Ca II NIR triplet in the majority of our spectra, and tentatively identify \ion{Ba}{2}, which had been found to be unusually strong in SN 1987A, in 9 events.
    \item The majority of the SNe II in our sample occur in environments with subsolar metallicities, in faint dwarf galaxies and the outskirts of large starforming galaxies. We find metallicities ranging from 0.08--0.92 $Z_\odot$ (using an indirect estimation method based on the host galaxy luminosity) or 0.29--1.08 $Z_\odot$ (directly measured from host galaxy spectra using the N2 emission line diagnostic). Two occur in near- or super-solar metallicity environments; this is consistent with the result from \citet{Taddia2013}'s metallicity study of SN 1987A-like SNe, which found two SNe in areas with near-solar metallicities. We also find one of the most metal-poor host environments known to date in SN 2019zfc, which has a directly measured metallicity of $12+\log{\textrm{(O/H)}} = 8.15$.
    \item Using our systematic sample of long-rising SNe II, we derive a robust lower limit to the local volumetric rate of these events, accounting for real survey conditions and measured incompleteness of galaxy redshift catalogs. We find the rate to be $1.4\pm0.3$\% of the total core-collapse rate in the local universe. This is consistent with previous estimates of the rate of SN 1987A-like events within a smaller volume.
\end{itemize}

Long-rising SNe II are observationally linked primarily by their unique dome-shaped light curves, whose shape is reasonably well-explained by a compact BSG progenitor \citep{Arnett1989,Kleiser2011,Taddia2013}. This is the basis upon which we systematically selected our CLU sample. Considered together, the long-rising SNe II in our sample and in the literature span a range of luminosities and colors, the strengths and velocities of important lines in their spectra vary, and even the tendency of these events to occur in low-metallicity environments has exceptions. This significant diversity makes modeling of BSG explosions difficult; for example, the single-progenitor modeling of \citet{Dessart2019} of previously known SN 1987A-like SNe revealed a heterogeneous class of SNe with no true prototype, and their models generally exhibited some level of discrepancy to observational data.

With our derived rate of $\approx$1.4\% of all CCSNe, we confirm that SN 1987A and similar SNe are indeed rare events. This rate is lower than that of \citet{Podsiadlowski1992}'s metallicity-independent estimate of $\sim$5\% of massive stars dying as BSGs due to mergers, so a simple merger scenario does not explain the landscape of BSG explosions. Similarly to the sample examined by \citet{Taddia2013}, in our sample a majority of SNe occurred in environments with subsolar metallicities. At the same time, both samples have several notable exceptions to this trend that occurred in environments with higher metallicities. It is likely that a both mergers and metallicity play a part in the BSG explosion mechanism, and the rarity of these events allows us to invoke more exotic progenitor scenarios.

Our addition of 12 new long-rising SNe II (from our sample of 13, as SN 2018hna has already been extensively studied) brings the total number of known SNe in this subclass within the local universe to $\approx$28. The increase in total sample size will enable more population-focused studies and improve our understanding of the full extent of the diversity within SNe with BSG progenitors. For example, the identification of any correlations between observational signatures, which requires a sufficiently large sample of events, could suggest further subclasses of BSG explosions that would, in turn, inform theoretical models and simulations. Early work in this area, notably \citet{Pumo2023}'s comparative study of 14 long-rising SNe II, has already challenged standard theories of stellar evolution and neutrino-driven core-collapse despite its small sample size.

Works such as this and \citet{De2020} have demonstrated that focused experiments for spectroscopic classification such as the volume-limited/volume-luminosity limited Census of the Local Universe and the flux-limited Bright Transient Survey yield meaningful samples of rare classes of transients from a wide-field, all-sky survey like ZTF. Volume-limited surveys such as CLU depend on currently incomplete catalogs of galaxies in our local universe, so will be greatly benefited by the improvement of the completeness of these catalogs through new spectroscopic surveys \citep[e.g., ][]{DESI2016,Kollmeier2017}. Furthermore, upcoming deep, wide-field time domain surveys, such as the Vera Rubin Observatory Legacy Survey of Space and Time \citep{Ivezic2019}, will enable us to further expand the known sample of BSG progenitor SNe to an even greater depth. In particular, in an era of large-scale surveys producing increasing volumes of photometric data that outpace limited resources for spectroscopic classification, the long rises and unique light curve shape of long-rising SNe II will make them relatively easy to identify through photometry alone. Combining the larger sample with more detailed host galaxy studies and identifying more individual progenitors will help shed light on the mysterious origins of SN 1987A-like supernovae.

\section*{Acknowledgements}
Based on observations obtained with the Samuel Oschin Telescope 48-inch and the 60-inch Telescope at the Palomar Observatory as part of the Zwicky Transient Facility project. ZTF is supported by the National Science Foundation under Grants
No. AST-1440341 and AST-2034437 and a collaboration including current partners Caltech, IPAC, the Weizmann Institute for Science, the Oskar Klein Center at Stockholm University, the University of Maryland, Deutsches Elektronen-Synchrotron and Humboldt University, the TANGO Consortium of Taiwan, the University of Wisconsin at Milwaukee, Trinity College Dublin, Lawrence Livermore National Laboratories, IN2P3, University of Warwick, Ruhr University Bochum, Northwestern University and former partners the University of Washington, Los Alamos National Laboratories, and Lawrence Berkeley National Laboratories. Operations are conducted by COO, IPAC, and UW. The SED Machine is based upon work supported by the National Science Foundation under Grant No. 1106171. 

The ZTF forced-photometry service was funded under the Heising-Simons Foundation grant \#12540303 (PI: Graham). The GROWTH Marshal was supported by the GROWTH project funded by the National Science Foundation under Grant No 1545949.

Some of the data presented in this work were obtained at the W.M. Keck Observatory, which is operated as a scientific partnership among the California Institute of Technology, the University of California and the National Aeronautics and Space Administration. The Observatory was made possible by the generous financial support of the W. M. Keck Foundation. The authors wish to recognize and acknowledge the very significant cultural role and reverence that the summit of Maunakea has always had within the indigenous Hawaiian community. We are most fortunate to have the opportunity to conduct observations from this mountain.

This research made use of the NASA/IPAC Extragalactic Database (NED) \citep{NEDcite}, which is funded by the National Aeronautics and Space Administration and operated by the California Institute of Technology, and Astropy,\footnote{\url{http://www.astropy.org}} a community-developed core Python package for Astronomy \citep{astropy2013, astropy2018}.

\bibliography{bibliography}
\bibliographystyle{aasjournal}

\end{document}